\documentclass[manuscript]{acmart}

\usepackage[utf8]{inputenc}
\usepackage{adjustbox}
\usepackage{array}
\usepackage{times}  
\usepackage{amsfonts}
\usepackage{amssymb}
\usepackage{wasysym}
\usepackage{hyperref}
\usepackage{listings}
\usepackage{verbatim}
\usepackage{hyperref}
\usepackage{booktabs}
\usepackage{colortbl}
\usepackage[shortcuts]{extdash}
\usepackage{subfig}
\usepackage{float}
\usepackage{graphics}
\usepackage{graphicx}
\usepackage{cprotect}

\definecolor{Gray}{rgb}{0.5, 0.5, 0.5}
\definecolor{ForestGreen}{rgb}{0.13, 0.55, 0.13}
\definecolor{red}{rgb}{1.0, 0.15, 0.07}

\newcommand{\good}[1]{\textcolor{ForestGreen}{#1}}
\newcommand{\mediocre}[1]{\textcolor{Gray}{#1}} 
\newcommand{\bad}[1]{\textcolor{red}{#1}}

\newcommand{\protocol}{PANEL}
\newcommand{\protocolfull}{\textbf{P}ractical \textbf{A}nonymity at the \textbf{NE}twork \textbf{L}evel}
\newcommand{\protocolfullnormalcap}{Practical Light-weight Anonymity at the Network Level}
\newcommand{\protocolsp}{\protocol \ }
\newcommand{\protocolfullsp}{\protocolfull \ }

\newcommand{\hcomment}[1]{}
\newcommand{\hedit}[1]{\textcolor{black}{{#1}}}

\newcolumntype{R}[2]{%
    >{\adjustbox{angle=#1,lap=\width-(#2)}\bgroup}%
    l%
    <{\egroup}%
}
\newcommand*\rot{\multicolumn{1}{R{55}{1em}}}

\setcopyright{rightsretained} 
\begin{document}

  \author{Hooman Mohajeri Moghaddam}
  \affiliation{ 
      \institution{Princeton University}
    }
    \email{ hoomanm@princeton.edu }
    
      \author{Arsalan Mosenia}
  \affiliation{ 
      \institution{Google}
    }
    \email{ mosenia@google.com }



  \title{Anonymizing Masses: \protocolfullnormalcap}



\begin{abstract}
{In an era of pervasive online surveillance, Internet users are in need 
of better anonymity solutions for online communications 
without sacrificing performance. Existing overlay anonymity tools, such as the Tor network, suffer from performance limitations and 
recent proposals to embed anonymity into Internet protocols face fundamental deployment challenges. In this paper, we introduce \protocolfullsp (\protocol), a practical light-weight anonymity solution based on \textit{hardware switching}. We implement a prototype of \protocolsp on a high-performance hardware switch (namely, Barefoot Tofino) using P4 network programming language, and examine the validity and performance of the prototype. Based on our empirical results, \protocolsp achieves 96\% of the actual throughput of the switch and adds a low-latency overhead (e.g., 3\% overhead in Skype calls), while offering partial deployablility and transparency (i.e., \protocolsp requires neither client-side nor server-side modifications). }
 
\end{abstract}



\maketitle

\section{Introduction}
\label{intro}
The state-of-the-art pervasive online surveillance apparatus is capable of recording, aggregating, and analyzing our online communications, allowing business and government agencies to control, monitor, and influence our online interactions~\cite{surv1,surv3}. Even when the communication content is encrypted, its \emph{metadata} (e.g., Internet addresses and protocol specifications) reveal critical information about the parties involved in the communication, which can be exploited by governments and corporations~\cite{nsa,surv}.

Unfortunately, popular anonymity systems today, such as Tor~\cite{tor}, have been only successful in attracting the most privacy-conscious users  due to their \textit{poor performance} \footnote{The total number of active Tor users is currently about 3 million users~\cite{tor-metric}, a figure that is far from the number of active Internet users, which is, by even conservative estimates, more than 3 billion users~\cite{interenetUsage}.}. Under the hood, legacy anonymity tools require dedicated infrastructure: a series of \textit{overlay proxies} to redirect traffic. This infrastructure typically consists of proxy servers run by volunteers, often with no performance guarantees, which causes bandwidth and latency issues~\cite{anonPerformance, torPerformance}.

To address the performance issues associated with the use of overlay approaches, a number of recent studies have proposed network-level anonymity solutions~\cite{torinsteadofip, lap, dovetail, hornet,  enlistingisp, phi,taranet}. However, these proposals are not readily applicable to today's Internet due to fundamental deployability challenges. Some solutions~\cite{hornet,torinsteadofip} \textit{entirely modify the Internet routing and forwarding protocols}, requiring a \emph{clean-slate} design approach~\cite{cleanslate}. For instance, they rely on end-to-end source-controlled routing model and its variants, such as segment routing~\cite{dovetail, hornet, torinsteadofip}, which are still experimental and not ubiquitously deployed~\cite{t0depr, ipv6srh}. 
Further, existing proposals lack a clear \emph{partial deployment} pathway~\cite{dovetail,hornet}, failing to take into account limitations imposed by different network actors in the wide area network, such as strict header format checking of Internet Protocol (IP) and transport layers by routers and middleboxes~\cite{ipOptionsNotOption,tcpext1}. 
Previous anonymity solutions ~\cite{lap, dovetail, hornet, enlistingisp, phi} are implemented and tested only on \emph{software switches}.

Our work is inspired by light-weight anonymity solutions (such as LAP~\cite{lap}); however, in this paper, we pursue an entirely different angle to anonymity systems, motivated by the observation that \textit{hardware switches} are pervasive in the Internet and offer a high performance and port density~\cite{ForwardingMetamorphosis}.  We present \protocolfullsp (\protocol), \textit{a partially-deployable, low-latency, light-weight} anonymity system compatible with both software and \textit{hardware switching}. In particular, \protocolsp offers: 

\begin{itemize}

\item \textbf{Network-level light-weight anonymity}: \protocolsp is implemented by Autonomous Systems (ASes) and embeds light-weight anonymity into the Internet forwarding infrastructure. This obviates the need for a dedicated overlay and can provide ubiquitous communication privacy to users by default. Further, it mitigates the undesirable scrutiny drawn by the opt-in nature of overlay systems~\cite{pew2}.

\item \textbf{Partial deployability and compatibility with legacy network:} \protocolsp  provides a \textit{practical approach for deployability}: it divides ASes on the communication path into \emph{\protocol-protected segments and legacy segments}. ASes in a single \protocolsp segment are at liberty to choose the anonymization method from existing proposals~\cite{torinsteadofip, lap, dovetail, hornet,  enlistingisp, phi}. However, on the boundary of a \protocolsp segment, designated border routers (referred to as \emph{landmark} routers) transform  packets to standard IP and transport format with fixed size headers, and forward these packets to the legacy network. Also, in \protocolsp the end-to-end path is decided by each hop individually (hop-by-hop routing).   
 
\item \textbf{Transparency:} As demonstrated in Section \ref{sec:eval}, \protocolsp is transparent to end hosts and does not require client-side (or server-side) modifications.

\end{itemize}

Our main contributions can be summarized as follows: 

\begin{enumerate}
\item We propose the first anonymity protocol designed for and tested on high-performance hardware switching fabric, offering low overhead and partial deployment. 

\item We propose a technique, that we refer to as \textit{privacy preserving line-rate source information hiding}. As discussed in 
Section \ref{source-info-rewriting}, this technique involves three main steps: (1) source address rewriting, (2) source information normalization (i.e., randomization of IP identification field and TCP initial sequence number), and (3) hiding path information (i.e., time-to-live randomization). Among these steps, source address rewriting based on software-switches has been mentioned in an earlier work \cite{enlistingisp}.  However, to the best of our knowledge, we are the first to implement privacy preserving source address rewriting based on hardware switches. 

\item We implement a prototype of the proposed protocol using a high-performance programmable hardware switch (Barefoot Tofino switch \cite{tofino}) and P4~\cite{p4} programming language.

\item We comprehensively evaluate our prototype and examine its validity and performance for different applications, such as web browsing and Skype video calls. Our empirical results suggest that that high-performance (line-rate) sender anonymity is achievable in today’s Internet, using high-performance hardware switches.

\end{enumerate}

\textit{This study sheds light on potential advantages of high-performance hardware switching for anonymity solutions, paving the way for further exploration in this area.} 

The rest of this paper is organized as follows. We discuss design goals in Section~\ref{sec:des-goals}, present an overview of \protocolsp in Section~\ref{sec:overview} and our design in Section~\ref{sec:sys-design}. The evaluation of our \protocolsp prototype appears in Section~\ref{sec:eval}, followed by a discussion of related work in Section~\ref{sec:related-work},  a discussion of limitations and potential extensions to \protocolsp in Section~\ref{sec:disc} and finally a conclusion in Section~\ref{sec:conc}.

\section{Design Goals}
\label{sec:des-goals}
In this section, we  outline the basic design goals of our protocol, including the privacy properties we aim to achieve. In Section~\ref{intro}, we highlighted partial deployment and compatibility with hardware switching fabrics as the two major factors towards practical deployment of network-level anonymity solutions. We address these issues in previous proposals and showcase a solution that is compatible with the legacy IP network, is designed to run on hardware switches, and achieves anonymity properties while being transparent to end hosts.

\subsection{Compatibility with Legacy IP Network}
\label{subsec:leg-compat}
We allow legacy and \protocolsp networks to coexist, enabling a partial deployment model. Here, we discuss desired properties that allow partial deployment of \protocol, which differentiates it from prior work~\cite{lap, dovetail, hornet, phi,taranet}.

\textbf{Standard Compliance:}
A number of prior proposals are based on packet-carrying state model~\cite{lap, hornet, phi}, in which session information is encoded in packet headers. However, carrying entire state information in packet headers proves to be impractical in the legacy network for the following reasons:

\begin{table}[]
\centering
\begin{tabular}{l|l}
     &  {Drop rate} \\ \hline
IPv4~\cite{ipOptionsNotOption, ipoptionnotoption2} & 30\% - 70\%                    \\ 
IPv6~\cite{ipv6-large-ext2} & 50\% - 90\%                    \\ 
\end{tabular}
\caption{IPv4 and IPv6 packet drop rate according to different measurement experiments~\cite{ipOptionsNotOption, ipoptionnotoption2, ipv6-large-ext2}.}
\label{tbl-drop-rate}
\end{table}

\begin{itemize}
\item First, while IPv4~\cite{ipv4rfc} and IPv6~\cite{ipv6spec} allow for IP options in the header, in practice support for these fields is very limited in today’s deployed networks. Table~\ref{tbl-drop-rate} summarizes the drop rate of packets containing IP options on the Internet, as reported by different measurement studies~\cite{ipOptionsNotOption, ipoptionnotoption2, ipv6-large-ext2}. 
The alternative approach of carrying  state in transport and application layer headers also faces deployment challenges: (1). Considering the transport layer,  the TCP options field is bounded in length (maximum of 40 bytes) and different protocols compete for that limited space~\cite{tcpext1}. (2). The fate of new experimental options are unknown due to pervasiveness of middleboxes enforcing transport and application layer policies~\cite{tracebox}. 
Also, embedding  routing state information into application layer space exposes packets to the risk of being dropped by network firewalls that monitor application layer semantics.
\item Second, packet-carrying state can leak information about the identity of the user, or the end-to-end route, unless routing information is encrypted. However, we avoid assuming that line-rate packet encryption is available at all landmark routers, for practical purposes.
\end{itemize}

\protocolsp aims to be compliant with protocol standards as well as the support offered by today’s deployed networks. Thus, \protocolsp \textit{does not use the packet-carrying state model and only adds a constant size small tag to the IP headers for packets sent to the legacy network.} 

\textbf{Routing Compatibility:} The source-controlled routing model, which many previous works are based on~\cite{dovetail,hornet}, provides a means to allow sessions sender to determine or influence the routing path.  However, source-controlled routing and its variants such as pathlet routing~\cite{pathlet}, are not ready for end-to-end deployment, due to unresolved incentive and security concerns~\cite{rfc5095}. Therefore,  \protocolsp operates in the \textit{hop-by-hop routing model}, where routing decisions are made independently by each landmark router. Also, \textit{\protocolsp allows for asymmetric routes between \protocolsp segments and for intra-segment routing}, as discussed in Section~\ref{sec:sys-design}.

\subsection{Hardware Switching}
\label{subsec:hard-switching}
The forwarding capacity and port density of hardware switching fabrics make them an indispensable part of the networking ecosystem from  data centers and service providers to large enterprise networks~\cite{ForwardingMetamorphosis}.
The advent of programmable switching fabrics~\cite{tofino,ozdag2012intel, xpliant,broadcom} and the abstractions provided by high-level programming languages supporting them~\cite{p4}, have enabled researchers to prototype and experiment with new protocols on hardware switching fabric for the first time. 

A network switch often consists of a switching chip that acts as the fast forwarding plane, often called data plane, and a slower control plane with a richer instruction set (e.g.  x86 CPU). Similar to OpenFlow~\cite{openflow} the  fast data plane can communicate with the control plane  via an API. 
Programmable data plane fabrics implement primitives such as \textit{match+action} tables and limited stateful packet manipulation, which are mapped to a high-level language specifications, such as P4. 
Our design challenge was to work within the constraints of the language and the specific switching fabric utilized to modify packets at line-rate, which we further discuss in Section~\ref{des:ses-unlink}.

\subsection{Privacy Properties}
\label{priv-definition}
The privacy properties discussed below form the basis of our   design. \hedit{We note that in our design we consider the trade-off between privacy guarantees and feasibility of deployment for our solution, thus our design falls into the category of light-weight anonymity solutions~\cite{lap}.}

\begin{itemize}
\item \textbf{Anonymity}:
A user, server, or a session is anonymous if it is not identifiable within an anonymity set, i.e., a set of entities with similar features~\cite{terminology}.  \hedit{In our light-weight anonymity model, we only provide \emph{sender anonymity}, which implies that the initiating party of a communication session must remain anonymous. }

\item \textbf{Unlinkability}:
The unlinkability property indicates that the chances of linking elements in the system (e.g., sessions and messages) is bounded by the a-priori knowledge of the observer, even after observing the system~\cite{terminology, systemsanoncommunication}. 
Given the capabilities of an adversary, if two sessions are unlinkable, no amount of observation from that adversary should increase his confidence that these two sessions are related.

\item \textbf{Path Information Protection}: Adding multiple layers of \textit{mixing} 
can increase the potential anonymity set, an idea first described by Chaum~\cite{chaum-untraceable} and one that 
has been used in multiple settings ~\cite{isdn-mixes, web-mix}. 
In our design, every additional \protocolsp segment adds a layer of mixing. However, topology-based analysis of packets with path information leakage can defeat the privacy benefits of iterative mixing (see Section 5.5 in \cite{hornet} for more detail). Therefore, \protocolsp ensures that packet headers do not leak path information, including path length.

\end{itemize}

\section{Overview and Environment}
\label{sec:overview}

 In this section, we first present a brief overview of \protocolsp design, then discuss the network environment and threat model.

\begin{figure}[h]
  \includegraphics[width=0.45\textwidth]{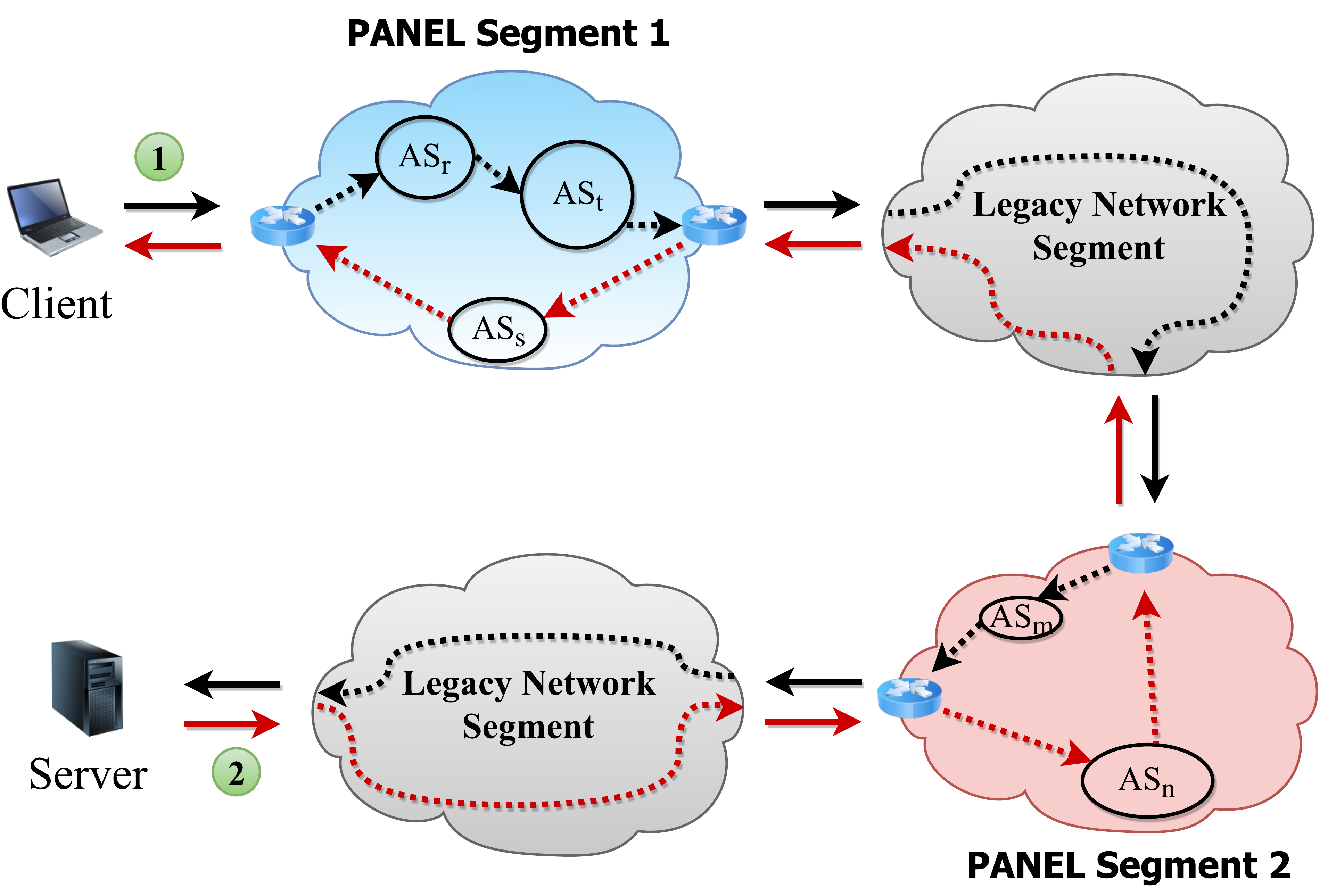}
  \caption{
  High level overview of \protocol: Packets for an end-to-end session from sender to receiver pass through a number of \protocolsp segments and legacy segments. Packet are rewritten at landmark  routers in both directions, from sender to receiver (Direction 1) and from destination to source (Direction 2). In each \protocolsp segment, routing and packet format are decided by the segment, thus asymmetric routing can be supported  internally and in legacy segments.}
  \label{fig:landmarks}
\end{figure}

\subsection{Overview}

A \protocolsp deployment is comprised of multiple \protocolsp segments, each of which contains one or more ASes. \protocolsp only dictates the behavior of landmark routers interfacing with the next segment and routing decisions internal to a segment are not defined by \protocol. Fig.~\ref{fig:landmarks} shows an end-to-end session from sender (client) to receiver (server).  In direction 1, packet headers are rewritten at the boundary of the \protocolsp segment to achieve anonymity and session unlinkability. Each landmark router rewrites the source addresses in packet headers and replaces them with local addresses. Session identifiers are also replaced with randomized identifiers, called \emph{tags}. The original address and session identifier, along with other session information (such as segment-specific information), are stored at landmark routers.
For the return path from the server to the client, each landmark, retrieves information stored locally and rewrites the headers in return packets in reverse direction. Thus, our work demonstrates two important ideas. First, that source information rewriting is sufficient to provide sender anonymity and session unlinkability in today's Internet. Second, \protocolsp achieves these properties at line-rate using existing abstractions available in programmable network devices with no client side modification.

As we will discuss in Section~\ref{routing-responses}, this approach forces the return route to uses the same set of landmark ASes in the reverse order and we will further discuss the implications for participating ASes in Section~\ref{impl-ases}. However, as shown in Fig.~\ref{fig:landmarks}, \protocolsp allows for asymmetric routes in the legacy network and  internally in each segment.

\subsection{Environment and Threat Model}
\label{threat-model}
An adversary's goal is to break the sender-anonymity or unlinkability properties of \protocolsp or reduce the anonymity set, as described in Section~\ref{priv-definition}.
Thus, \protocolsp provides these privacy properties against session receiver, every Autonomous System (AS) and passive or active eavesdroppers  that are capable of observing, dropping, replaying and modifying packets either on the path or in compromised routers.
We allow adversaries with a partial view of the network and those which are only capable of observing and controlling only a portion of ASes on the path. Similar to real-time anonymity systems such as Tor~\cite{tor}, global adversaries and end-to-end timing analysis are outside the scope of \protocol~\cite{murdochTorTrafficAnalysis,hopper2010much}.

To ensure \protocolsp is transparent to end-hosts, we delegate the task of protecting user privacy to the Autonomous Systems. \protocolsp provides privacy benefits even with a single AS deployment, however, additional \protocolsp ASes on an end-to-end path increase the anonymity set and improve session privacy. In the absence of other network-level anonymity solutions such as Dovetail~\cite{dovetail} or Phi~\cite{phi}, this means that the first \protocol-AS must be trusted, reducing our adversary model 
to that of the light-weight anonymity protocols~\cite{lap}. 
We assume the interface between a \protocolsp segment and legacy network is defined by Internet Protocol (IPv4~\cite{ipv4rfc} or IPv6~\cite{ipv6spec}). 

Finally, while payload encryption and more advance privacy preserving mechanisms such as onion routing~\cite{onion-routing} can help improve the confidentiality of content and metadata, we do not require that \protocolsp landmark routers support encryption at line-rate. Thus, \protocolsp leaves these functionalities to application and transport layers~\cite{tor, tcpcrypt, halfTrafficEncrypted}, in order not to overburden \protocolsp routers with cryptographic requirements, simplifying the deployment of \protocolsp on existing routers that do not support such capabilities at line-rate. 
\section{\protocolsp Design}
\label{sec:sys-design}
In this section, we explain how we achieve anonymity and session unlinkability on programmable switches. Our design is heavily influenced by our objective to make \protocolsp practical and deployable in today's Internet and applicability to primitives on programmable hardware switching fabric, such as match-action tables~\cite{ForwardingMetamorphosis}. Also, we demonstrate how our design is compatible with legacy networks both in terms of routing and packet header format.

\subsection{Privacy Preserving Line-rate Source Information Hiding}
\label{source-info-rewriting}
This section describes how to achieve anonymity through in-network source information hiding, based on the primitives available in programmable switching chips.

\subsubsection{Source Address Rewriting}

\label{source-anonymity-rewriting}

\begin{figure}
  \includegraphics[width=0.5\textwidth]{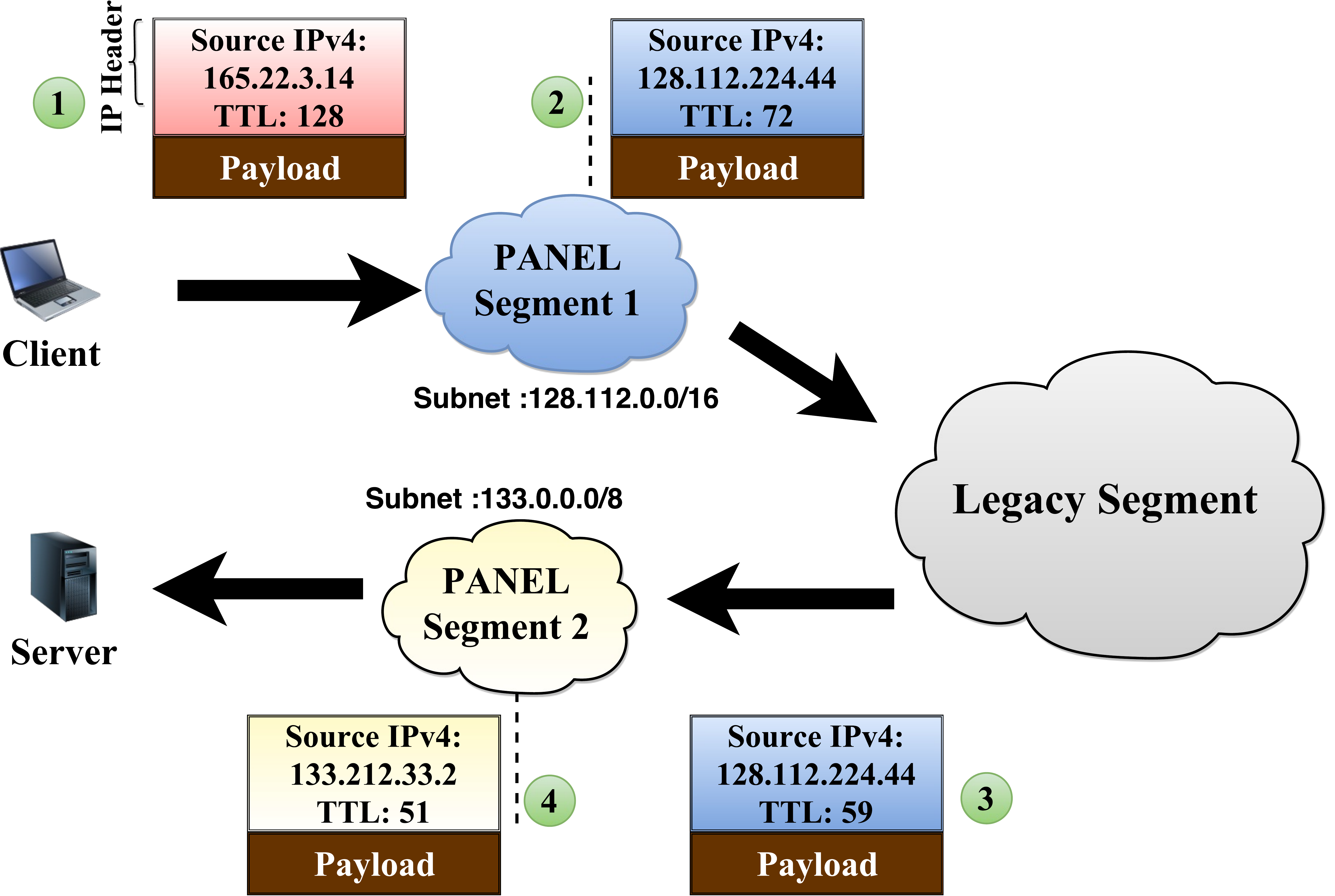}
  \caption{Source information rewriting in IPv4 network: a packet originating from the sender (1) is transformed in the first \protocolsp segment (2), with its source address and TTL randomized. The source IP address chosen at each landmark is from the \protocolsp IP pool. The packet is then forwarded in the legacy segment (3) as a standard IP packet. Additional \protocolsp segments (4) further anonymize the packet source by replacing its source address.}
  \label{fig:src-addr}
\end{figure}

While the idea of source address rewriting  is not novel, we are the first to demonstrate how to achieve privacy preserving source address rewriting at line-rate on hardware switching fabric,  without relying on overlay proxies. Thus, a major contribution of  \protocolsp is to  demonstrate that sender anonymity through \emph{source rewriting} is achievable in today's Internet, using high performance switching chips.

Fig.~\ref{fig:src-addr} shows how the concealment the origin of packets is made possible through source information rewriting on a per-session basis. First, each landmark AS  sets aside a subnet of public IP addresses, called \emph{\protocolsp (IP address) pool}, which will be advertised through BGP and will be used to replace the source IP addresses for all sessions.
A session on route from source to destination is modified at each landmark \protocolsp router, replacing the source address with an IP address from the \protocolsp pool. As we will see in Section~\ref{tag-generation}, the IP address is chosen at random using a Pseudo Random Number Generator (PRNG). 
In Fig.~\ref{fig:src-addr}, an instantiation of source rewriting is shown based on IPv4 addressing. For each packet, the source address is rewritten at each landmark router and packet is sent to the legacy network where the routing and forwarding take place without any changes in the legacy network.

\subsubsection{Source Information Normalization}

\begin{figure}
\centering
 \includegraphics[width=0.48\textwidth]{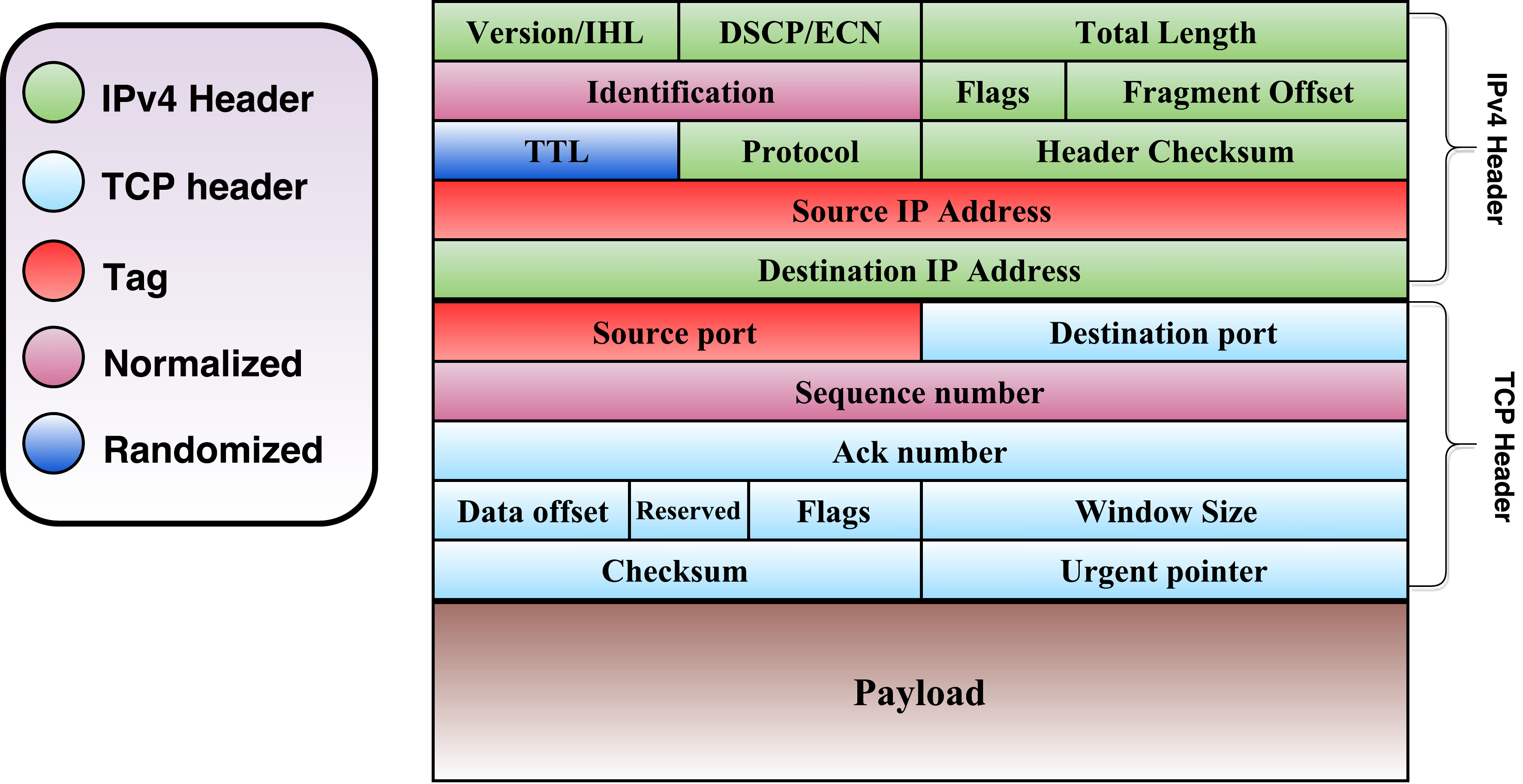}
  \caption{Modified TCP/IP header: \protocolsp modifies headers in IP and transport layers (TCP). Source IP address and port number fields are used to place randomized tags. IPv4 identification field and TCP sequence numbers are normalized to thwart passive fingerprinting attacks. Finally, TTL (Hop Limit) field is  randomized.}
  \label{fig:tcp_ip_hdrs}
\end{figure}

Several fields in packet headers can be utilized to perform fingerprinting attacks. In the presence of passive adversaries, IP identification field and TCP initial sequence number (shown in Fig.~\ref{fig:tcp_ip_hdrs}) reveal information about clients' software. These fields are implemented using different algorithms on popular operating systems~\cite{linux-tcp-isn, windows-tcp-isn, openbsd-tcp-isn} and exploiting these fields to identify clients or to link different sessions of a user is a well known technique~\cite{bellovin2002technique, tcp_ip_classifier}.  Thus, the first landmark router normalizes these fields in packets to minimize the chances of such attacks. The landmark router must ensure that it is the first \protocolsp segment on the path between sender to receivers\footnote{If the packet is not originated in the same segment as the landmark router, the router can check whether the packet's source address belongs to a \protocolsp segment, by consulting a global list of IP addresses of \protocolsp segments.}. Upon establishment of a session on a landmark router, a random offset is generated for each of the fields that will be added in one direction and subtracted in the reverse direction.

\subsubsection{Hiding Path Information}
\label{hide_path_info}

In the legacy IP network, time to live (TTL) or hop limit field in packet headers also leaks information about the distance to the original sender at any given vantage point. Assume that the initial TTL value $T_i$ is a fixed constant $c$. If an adversary observes a specific TTL value $\delta$, then he can readily calculate the hop distance ($d$) to the packet sender. Furthermore, through pre-computation on the AS-level topology, the adversary can narrow down the list of ASes where this packet could have originated from. This can be detrimental  to the iterative mixing property of \protocol, discussed in Section~\ref{priv-definition}. Therefore, the initial TTL value $T_i$ must be randomized by a landmark router, as shown in Fig.~\ref{fig:tcp_ip_hdrs}. To determine the probability distribution for TTL values, we use mutual information to quantify the information leakage for a distance $d$, through observing value $\delta$. We assume $d$ is chosen from a distribution $D$ and our objective is to find a distribution $\Delta$, such that $T_i$'s chosen from this distribution leak the least information about the distance to source at any given vantage point. Thus, our objective function is the following:

\begin{equation} \label{min_delta_eq}
\begin{aligned}
& \underset{\Delta}{\text{minimize}}
& & I (\Delta  - D; \Delta) \\
& \text{subject to}
& & supp(\Delta) = \;\{  T_{min}, \ldots, T_{max}\}.
\end{aligned}
\end{equation}

Where $I$ is the mutual information function, $supp(\Delta_i)$ is the range of possible values for $T_i$'s and $T_{min}$ and $T_{max}$ are minimum and maximum values that the initial TTL value can take. Note that, by definition, the initial TTL has to have an upper bound ($T_{max}$). Also, it cannot be set to zero and has to be large enough such that packets on long end-to-end paths do not expire, thus a positive $T_{min}$ is needed (e.g. $64$). We will discuss in Section~\ref{eval:ttl} how to empirically estimate $D$ and how to compute the distribution  $\Delta$.

\subsection{Session Unlinkability with Randomization}
\label{des:ses-unlink}

\begin{figure}
\centering
  \includegraphics[width=0.325\textwidth]{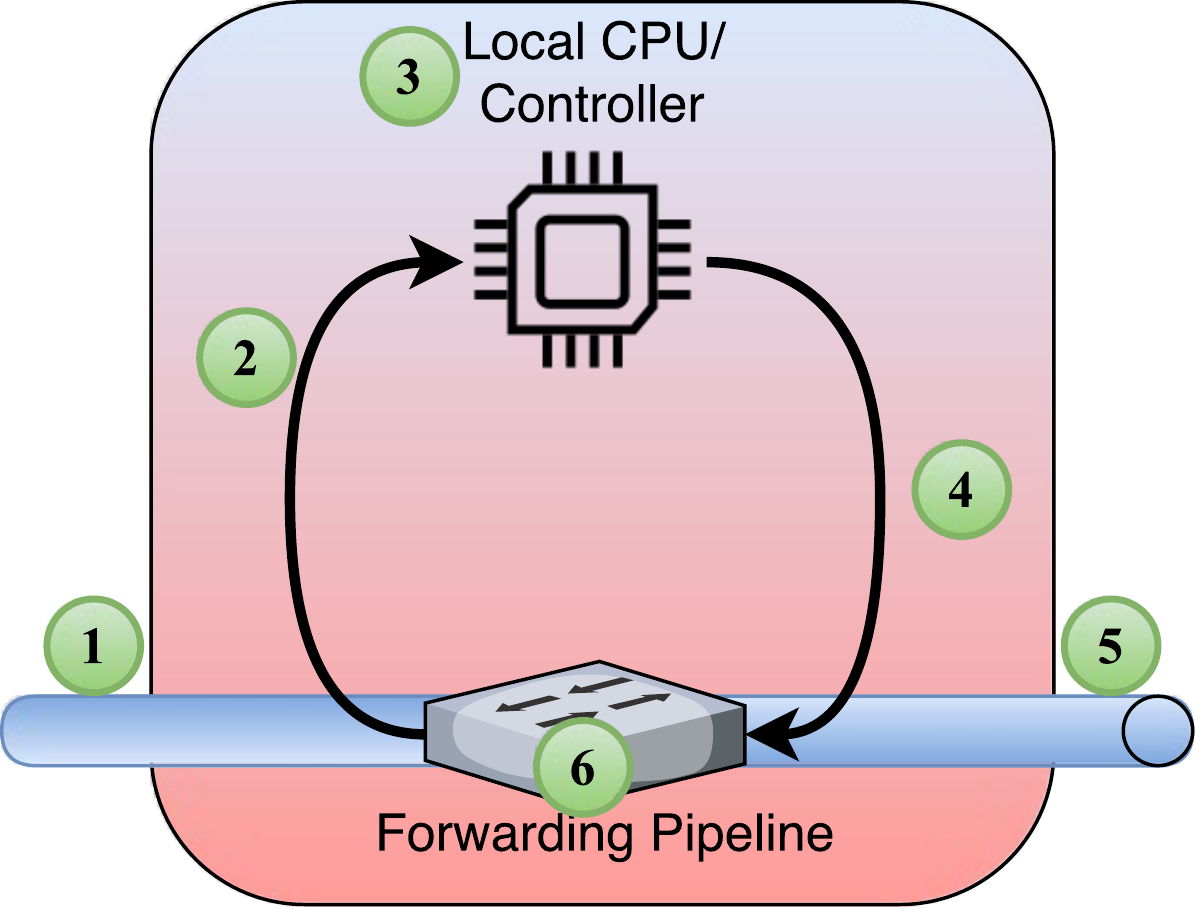}
  \caption{Once the first packet for a session arrives at the switch (1), the packet is sent to a local CPU or a specialized chip (2), where a random tag is generated for that session (3). The tag along with session information are inserted into the forward table (4). Finally, the packet is modified and forwarded to the destination (5). Subsequent packets for this session are matched and modified in the forward table, stored in the switch data plane (6).}
  \label{fig:switch-chip-internal}
\end{figure}

\begin{figure}
  \centering 
  \includegraphics[width=0.49\textwidth]{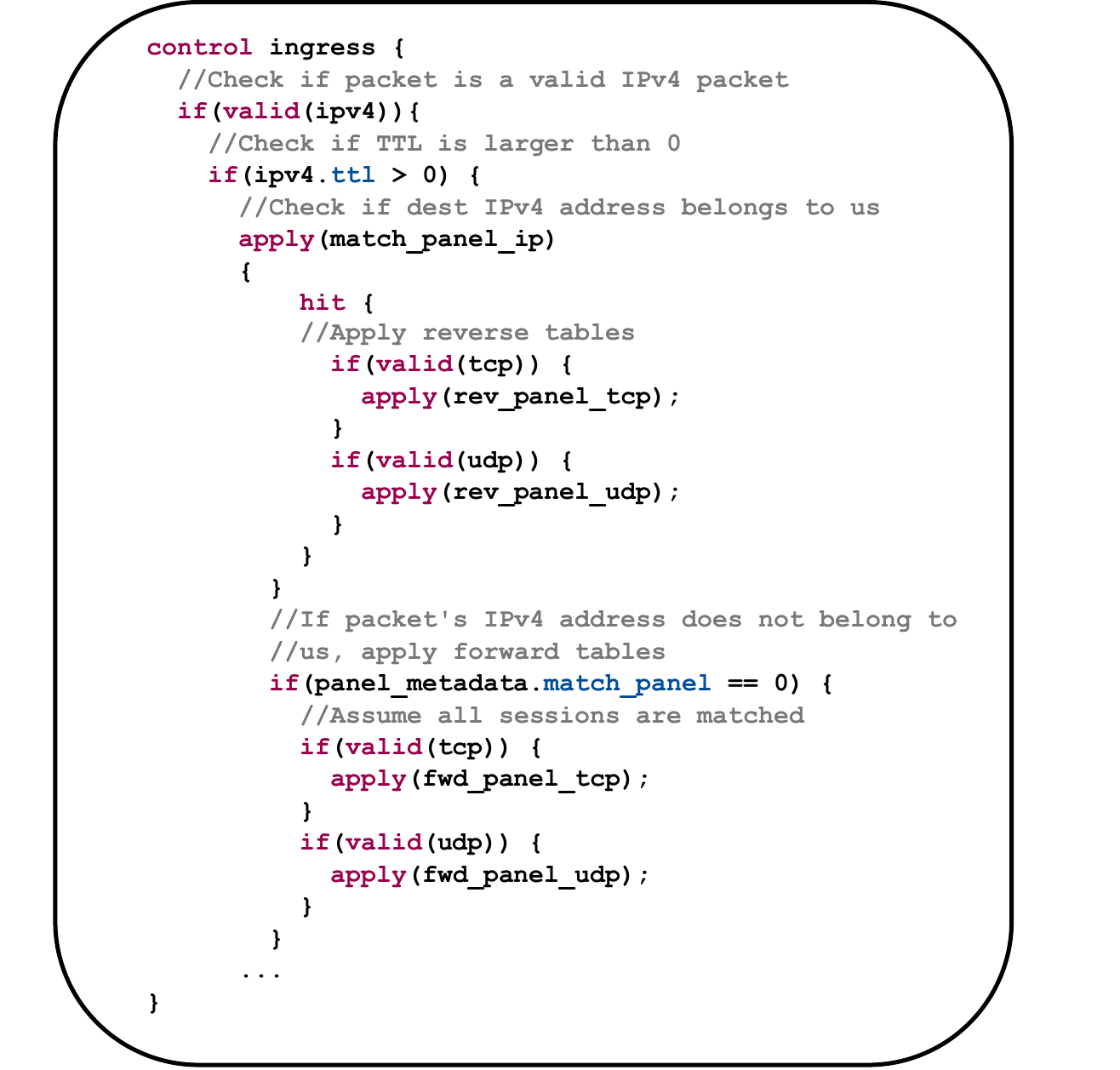}
  \cprotect\caption{Ingress pipline of \protocolsp written in P4 for IPv4 addresses. Packets matching an IP address that belongs to the switch, will be handled by reverse tables, i.e. \verb|rev_panel_tcp| or \verb|rev_panel_udp|. Other packets are handled by forward tables, i.e. \verb|fwd_panel_tcp| or \verb|fwd_panel_udp|.}
  \label{fig:p4-code}
\end{figure}

To achieve session unlinkability, session identifiers, such as TCP ports, are  randomized in \protocol. There are two challenges with randomizing session identifiers. 
First session identifiers must fit into small fixed size header space to be compatible with legacy network.
Second, we do not assume that per session asymmetric cryptography is available at every landmark router, so routers must assign session identifiers independently. 
Therefore, we replace session identifiers with the output of a cryptographic pseudorandom number generator (PRNG), called ``\emph{tags}".
 As shown in Fig.~\ref{fig:switch-chip-internal}, we assume that a local agent or an external chip on the router generates these random tags for each new incoming session and guarantees that active tags are not reused. The tag is split and stored in lower bits of the source address and transport layer port numbers and further parts of IP and transport headers, when needed (discussed further in Section~\ref{tag-generation}). 
 Using secure random number generation guarantees that packet tags are independent of each other and  provide tag bitwise unlinkability~\cite{george-thesis}. Once a tag is assigned to a session, the tag and the corresponding source information is stored in a key-value table, referred to as \emph{forward table} and each subsequent packet will be matched by its source information (key)  to retrieve its corresponding tag (value), as shown in Fig.~\ref{fig:p4-code}.
 Then the packet will be rewritten similar to the first packet using a match-action functionality available in programmable switching fabric~\cite{ForwardingMetamorphosis}. As we will discuss in Section~\ref{sec:valid-perf}, in our current implementation key sizes are reduced using a hash function in the data plane.

\subsection{Forwarding Responses in Absence of Return Addresses}
\label{routing-responses}

Since routing on the Internet is based on the destination addresses, removing original source addresses at landmark nodes makes routing response packets challenging. Here, we describe how our design overcomes this challenge. 

\begin{figure}
  \includegraphics[width=0.48\textwidth]{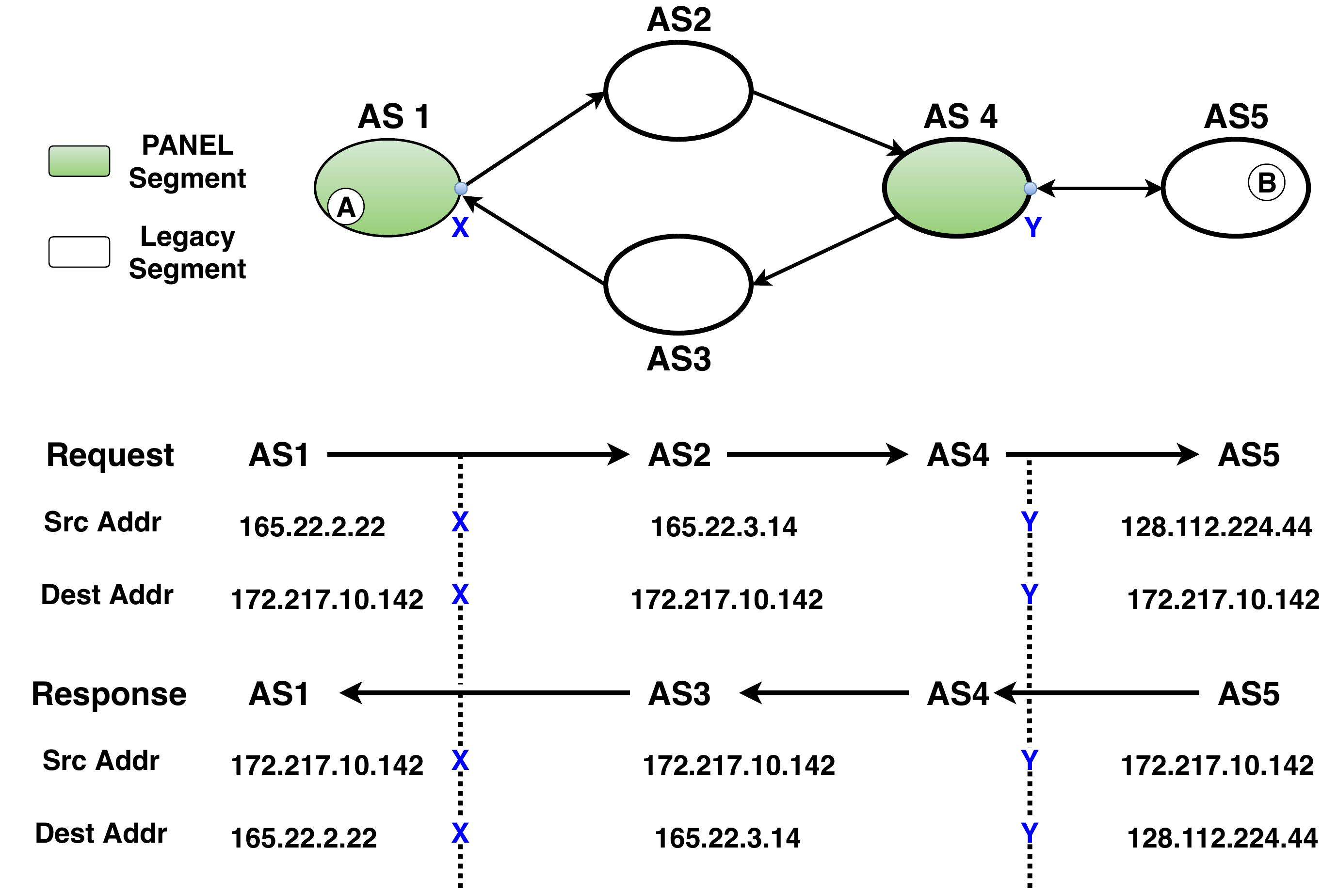}
  \caption{When $B$ sends a response to $A$, the return address of $A$ is absent in the packet header. $B$ knows that the request  packet passed through $Y$ because of the source address  of the request packet observed by $B$ belongs to $Y$. Thus, $B$ will send its response to $Y$. $Y$ receives this packet addressed to its landmark IP, checks its list of tags, and sees that the request originated from $X$.  When the response arrives at $X$, addressed to its landmark IP, $X$ checks its list of tags, and sees that the request was sent by $A$. Finally, the response has its destination IP address modified to the IP address of $A$ and forwarded from $X$ to its final destination, $A$.}
  \label{fig:flows}
\end{figure}

To be able to route the response, a landmark router must store the original session information.
This information is stored in another table, referred to as \emph{reverse table}. Reverse table is a look-up table that conceptually applies the reverse actions of forward table. Fig. \ref{fig:flows} illustrates this translation in  an asymmetric routing scenario.

\subsubsection{Match-Action Translation}
\label{tag-generation}
\hcomment{Do I need a figure for match action table?}

\begin{figure}
  \includegraphics[width=0.5\textwidth]{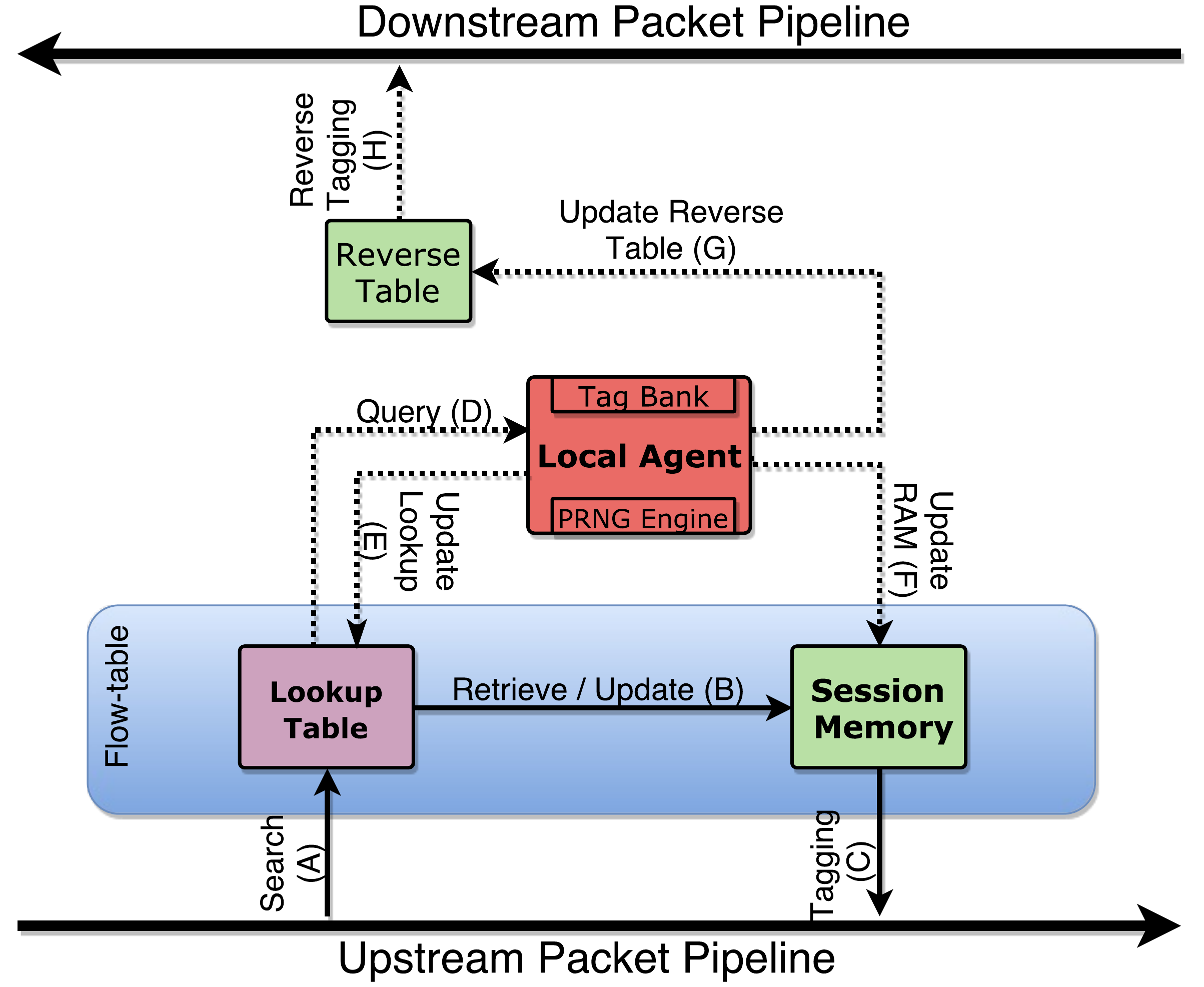}
  \caption{  \protocolsp packet processing pipeline: if a session already exists in the forward table, the packet source information is searched in the table (A) and corresponding tag is retrieved from the memory (B) and is applied to the packet header(C). If, however, the session is not in the table, on the first packet of the flow, the local agent is queried (D) for a new tag. Once a free tag is assigned to the session, the local agent updates the tables (E and F) to reflect the changes. For downstream packets another match-action table, named \emph{reverse table} is updated by the local agent to perform the reverse tagging operation, which performs the operation described in Section~\ref{routing-responses}. }
  \label{fig:tag-gen}
\end{figure}

Thus far, we have established that in \protocol, fixed size tags are placed in packet headers at every landmark node to identify sessions and sources in the absence of original headers and these tags are placed in the TCP/UDP port numbers and lower bits of IP address field.
We apply tags to sessions through match-action tables~\cite{ForwardingMetamorphosis}, referred to as {forward and reverse
 tables} in the data plane of the router, as shown in Fig.~\ref{fig:tag-gen}. Upstream\footnote{ We refer to sessions from the source to the destination as \emph{upstream flows} and the return sessions as \emph{downstream flows}.} packets are tagged using a lookup table. The first packet of each session will hit the local agent or a customized chip for tag generation, where tag is a unique identifier generated from a PRNG.

\subsection{Composition with other Anonymity Systems}
\label{sec:composition}
\begin{figure}[h]
  \includegraphics[width=0.48\textwidth]{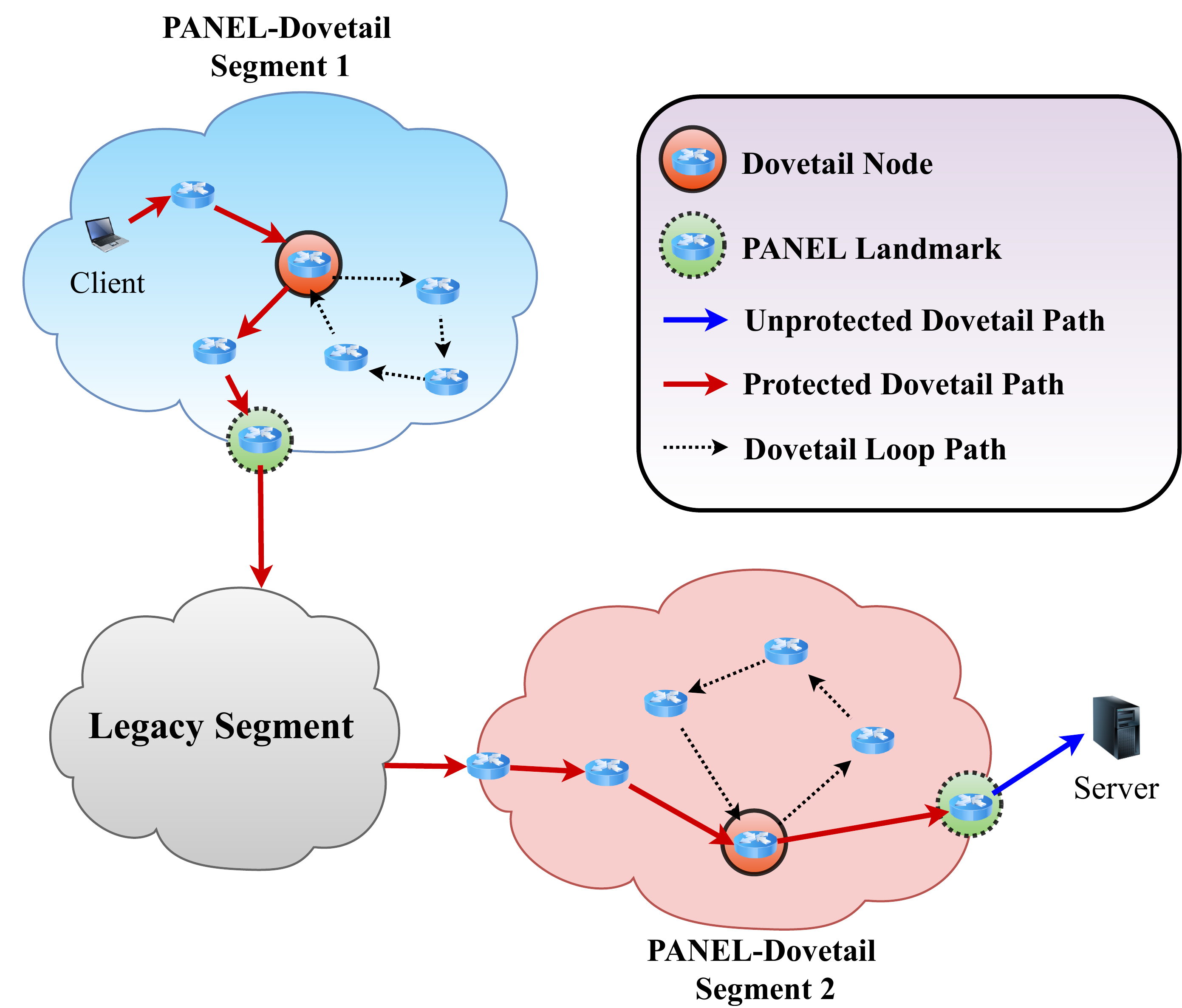}
  \caption{\protocol-Dovetail: \protocol-Dovetail allows Dovetail sessions to be extended over multiple segments, with the first segment establishing and end-to-end path from sender to the last \protocol-Dovetail segment landmark router. Thus, the last landmark router  acts as a transparent proxy for the client, decapsulating  \protocol-Dovetail layers, similar to a Tor exit.}
  \label{fig:panel-dovetail}
\end{figure}

\hedit{
Recall that PANEL provides only light-weight anonymity by focusing on sender anonymity, and omits receiver anonymity and payload encryption from its design. 
By composing \protocolsp with other anonymity systems, we can augment the privacy or performance properties of these solutions. For instance, Dovetail (discussed in Section~\ref{sec:related-work}) is a network-level anonymity mechanism that provides receiver anonymity, thus, \protocol-Dovetail enables sender/receiver anonymity.} Another example is that of the Tor network, which  is vulnerable to traffic analysis attacks if both the Tor guard and the Tor exit are malicious and colluding. The composition of \protocolsp with Tor helps mitigate such traffic analysis attacks by protecting the identity of clients from Tor guards. 
Below we showcase  \protocol-Dovetail and \protocol-Tor:

\textbf{\protocol-Dovetail}: \hedit{One of the advantages of the Tor network is that the end receiver for a Tor session does not have to be part of the Tor overlay network. However, both Dovetail and HORNET require both the session sender and the receiver to be part of the segment. As shown in Fig.~\ref{fig:panel-dovetail}, \protocol-Dovetail allows for a receiver outside the last segment. Anonymity headers are added in the first segment and removed by the last segment similar to MPLS~\cite{mplsrfc}, enabling a client to  transparently access an end host through a  \protocol-Dovetail deployment.}

\begin{figure}
  \centering
  \includegraphics[width=0.4\textwidth]{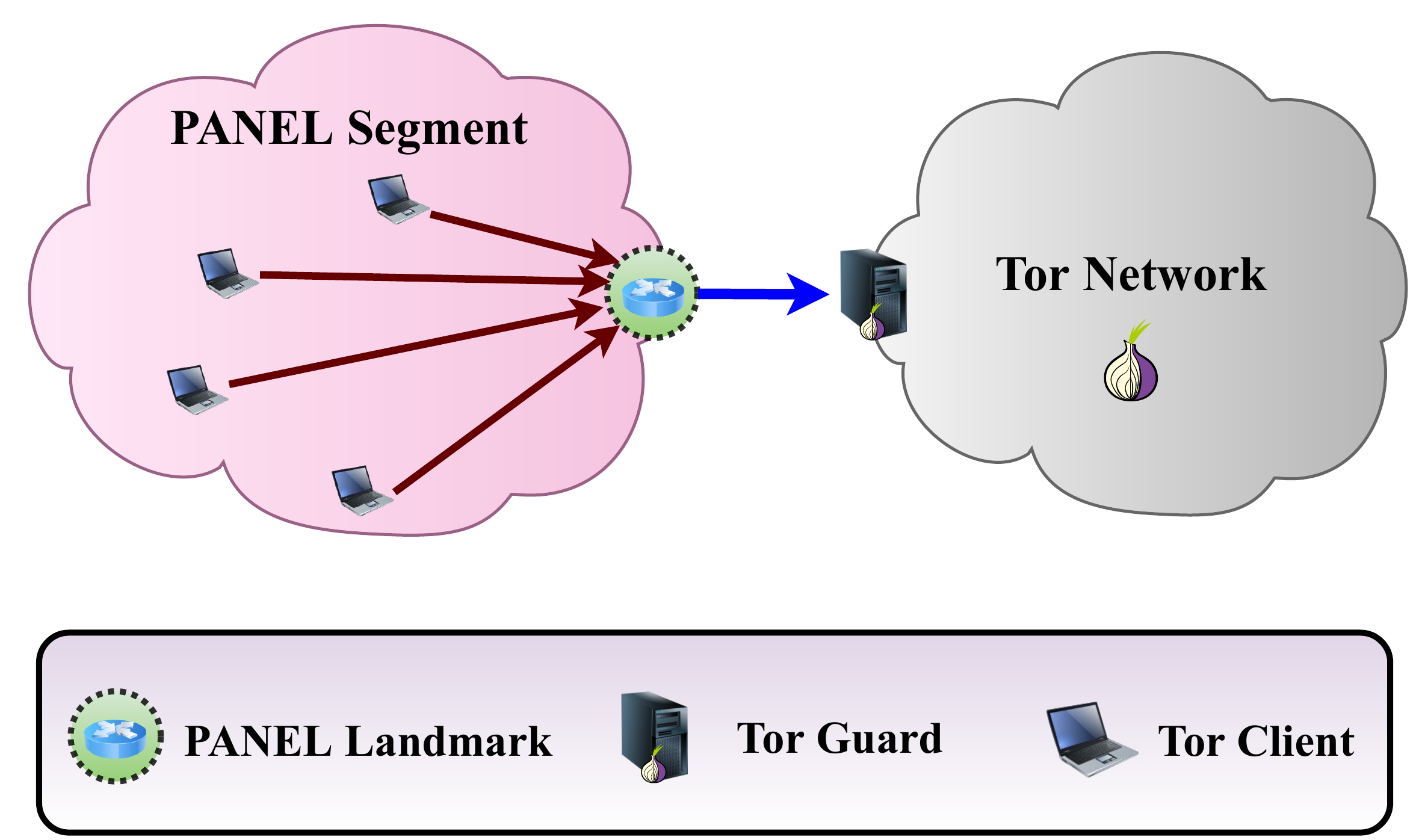}
  \caption{\protocol-Tor: A \protocolsp segment can protect the users connecting to the Tor network by hiding their identity}
  \label{fig:tor-panel}
\end{figure}

\textbf{\protocol-Tor}: \hedit{The Tor anonymity network is still popular, despite the performance issues it is facing. A key challenge in the Tor network is how to securely select the entry node (guard) for each client connecting to the network~\cite{dingledine2014one}. Since Tor's guard selection algorithm takes guards' available bandwidth as an input, powerful adversaries can inject high bandwidth guard nodes (and exit nodes) to compromise client anonymity. 
For example, such attackers can perform website fingerprinting attacks or even end-to-end timing analysis attacks, if they can monitor or compromise the corresponding Tor exit. With \protocol-Tor, however, guard nodes are no longer able to identify the client connecting to them, as shown in Fig.~\ref{fig:tor-panel}.}

\section{Evaluation}
\label{sec:eval}
In this section, we evaluate the feasibility of our approach on programmable switches. We demonstrate our prototype followed by an empirical study of the parameters in our design and an estimate of the anonymity size of our solution in today's Internet.

\subsection{Hardware Implementation}

\hcomment{Mention how we only add table entries on a TCP SYN\/ACK  and how we timeout sessions.}

Our implementation is based on a real-world  Barefoot Tofino switch~\cite{tofino} with 3.3 Tbps forwarding capacity. \protocol's forwarding plane code is written in a) P4, which is a field reconfigurable, protocol and target independence programming language designed as a platform for programming parse-match-action pipelines,
through match-action tables in hardware, and b) a local agent written in python, running on an Intel x86 CPU, which performs tag generation.

\begin{figure}
  \centering 
  \includegraphics[trim={1mm 1mm 3mm 2mm},clip,width=0.49\textwidth]{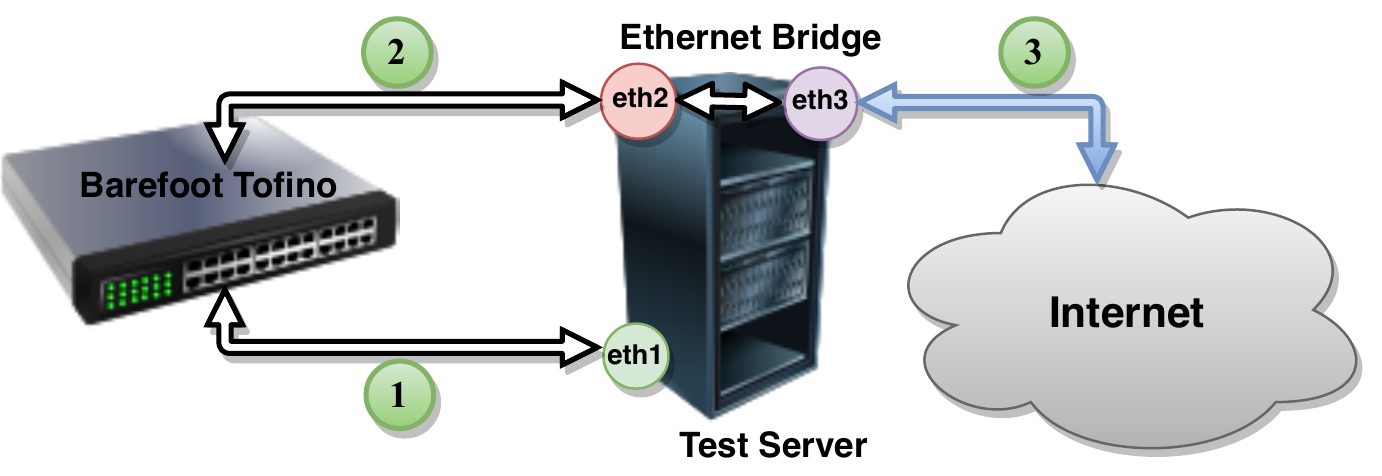}
  \caption{Experiment setup: Our test server is equipped with 3 Ethernet interfaces, each isolated to a single Linux namespace. $Eth1$ and $Eth2$ are connected to the Tofino switch over a dual port network interface card. All traffic received on $Eth2$ is forwarded to $Eth3$ interface that is connected to the Internet over a local Ethernet bridge.}
  \label{fig:tofino-exp}
\end{figure}

Our testbed consists of a server connected to the Tofino switch through a Mellanox ConnectX-3  dual-port 40GbE network interface card. As shown in Fig.~\ref{fig:tofino-exp}, each of the interfaces connected to the switch are isolated to a Linux namespace. Packets are passed through the switch via one Ethernet interface and received on the other interface. Finally, each packet is forwarded to its destination on the Internet via a local Ethernet bridge.

\subsection{Validation and Performance}
\label{sec:valid-perf}
\begin{enumerate}

     \item \textbf{Validation}: In order to demonstrate that \protocolsp runs transparent to clients and application-level protocols, we ran two sets of experiments on our testbed using standard Ubuntu 16.04. First, we tested DNS, Web and TLS protocols by fetching 100 different domains from Alexa top websites over both HTTP and HTTPS
     . Each page was fetched  with \protocolsp and vanilla routing software (\emph{simple router}) and inspected visually to make sure the resulting pages are identical. Also, we successfully accessed servers over a range of protocols, including FTP, SSH, POP3, SMTP, Telnet and SMTP both with and without \protocol. Next, for real-time applications we tested Skype video calls. We verified  that calls are established and can last over 10 minutes.

\item \textbf{Throughput}: Using iPerf{\footnote{\url{https://iperf.fr}}}, we first measured the bandwidth of packet forwarding between the two interfaces connected to the switch. These interfaces are hosted on the same network interface card that runs at 20 Gbps on each port. The average bitrate over one minute experiments were 16.9 $\pm$ 0.38 Gbps and 16.2 $\pm$ 0.54 Gbps for simple router and \protocolsp respectively, showing only a 5$\%$ bandwidth overhead. We repeated the same experiment with a remote iPerf server, using the Ethernet bridge depicted in Fig.~\ref{fig:tofino-exp}. The bandwidth limit in this case is dictated by our upstream link to the Internet. For this experiment, simple router and \protocolsp respectively yield 127 $\pm$ 14 Mbps and 118 $\pm$ 22 Mbps bitrates. \emph{These experiments show that we can achieve close to 92-96$\%$ of the switch capacity.} Thus we expect that if we connect all the 32 ports on the Tofino switch to 100 Gbps interfaces, we can achieve Tbps forwarding capacity.

\item \textbf{Latency}: We measured the latency for the Skype experiment above. 
We repeated the Skype experiment 10 times and measured the latency in the video received. As shown in Table~\ref{tbl-skype-latency},  \protocolsp adds only $3\%$ latency over simple router for Skype calls. We also employed another method to measure latency incurred by \protocol. Using Gnu Wget software we fetched 100 websites from Alexa\footnote{\url{https://www.alexa.com/topsites}}, 100 times each. 
Fig.~\ref{fig:wget-latency} compares the latency of \protocolsp and simple router for each of these 100 websites. We discuss how the majority of this latency occurs during the first round trip in  Appendix~\ref{appA}.
\begin{figure*}[]
   \centering 
	    \includegraphics[trim={10mm 1mm 10mm 5mm},clip,width=450px, height=280px]{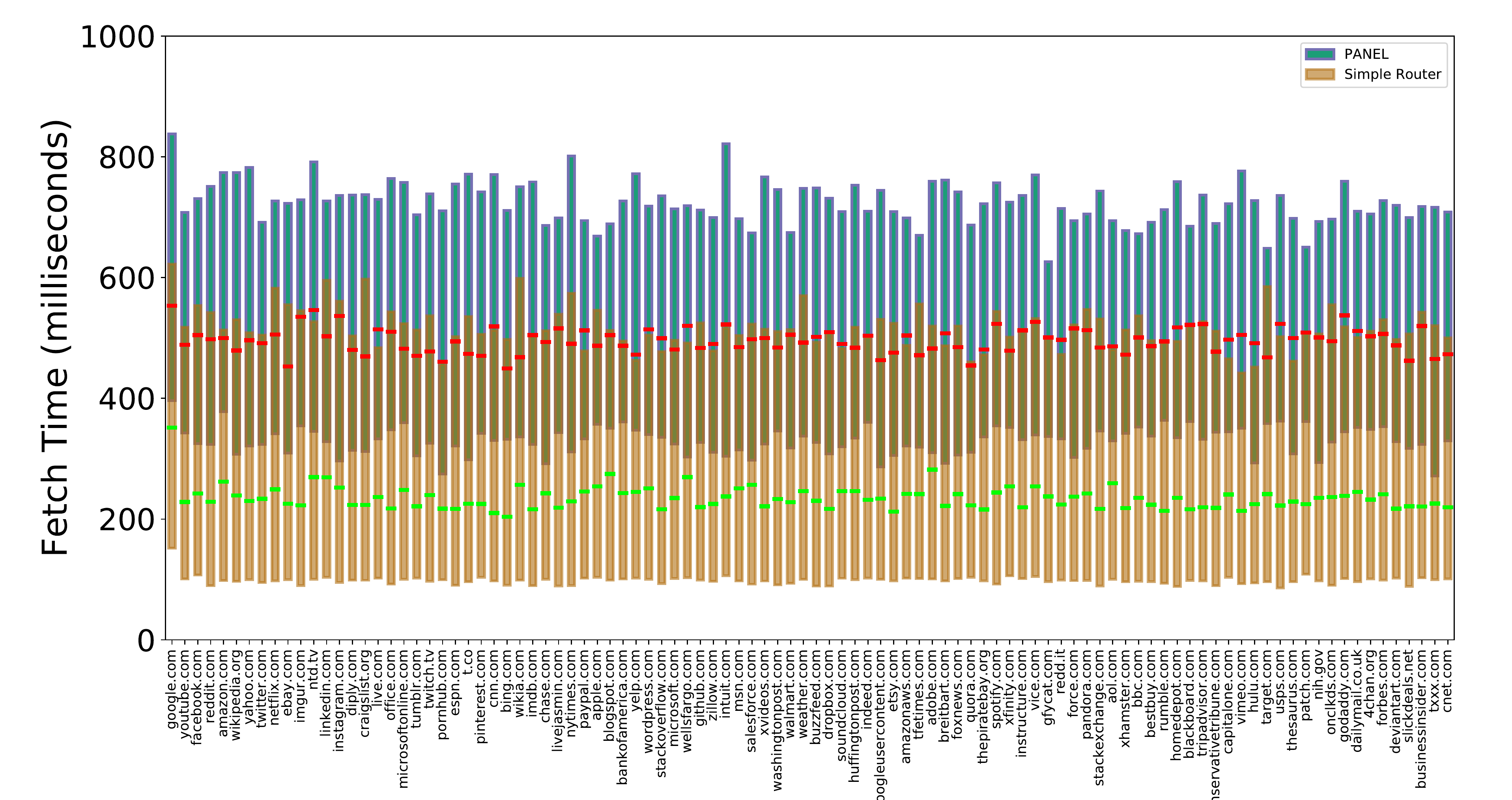}
  \caption{Comparison of latency for Alexa top 100 domain over \protocolsp and simple router. This figure shows the latency to fetch the front page of each domain.}
      	\label{fig:wget-latency}
\end{figure*}

\begin{table}[]
\centering
\begin{tabular}{c|cc}
     &  {Latency (millisecond)} \\ \hline
Simple router & $ 241.33 \pm 31.14$ \\ 
\protocol & $248 \pm 52.40 $\\ 
\end{tabular}
\caption{Skype video call latency in Simple router vs. \protocol.}
\label{tbl-skype-latency}
\end{table}

 \item \textbf{Switch Capacity}: To understand the number of sessions our Tofino switch can keep in memory, match-action tables with different key length were compiled on the switch, as shown in Fig.~\ref{fig:tofino-cap}. If the switch had unlimited memory, the (theoretical) bound for table sizes would grow exponentially with the key length, however, in practice the switch reaches the maximum capacity for keeping the entire table in the memory at key length of 21 bits. Therefore, we use a hash function in the data plane to truncate the tag to match this key length.

 \begin{figure}
  \centering 
  \includegraphics[width=0.49\textwidth]{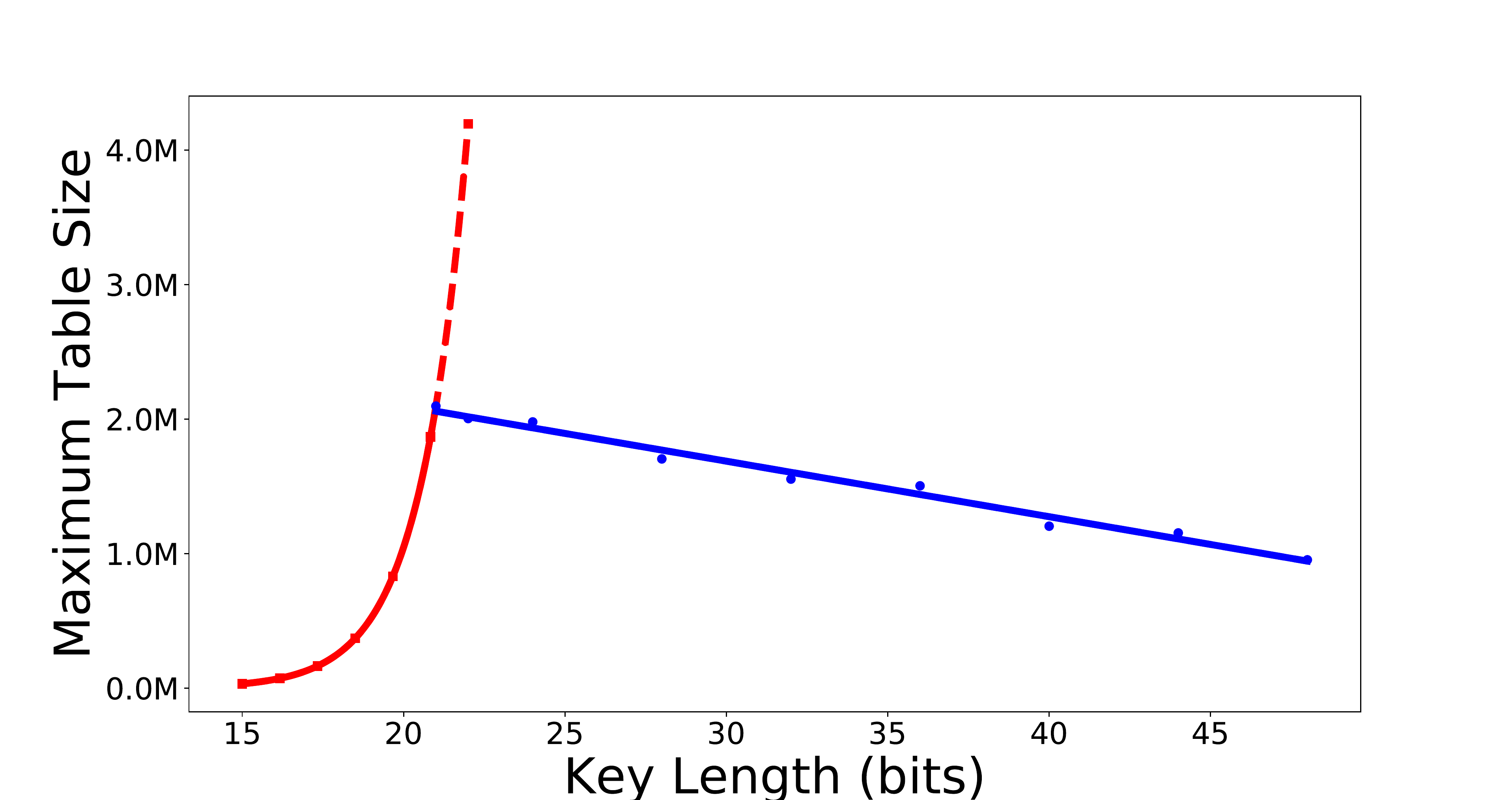}
  \caption{Tofino switch capacity: Table sizes with unlimited memory (\textit{theoretical bound}) shown  in red and the maximum table size allowed on the Tofino switch (\textit{practical bound}) for each key size, shown in blue. For key lengths less than or equal to 21, the theoretical bound matches the practical bound, but after the 21 bit key length, we reach the SRAM memory limit on the switch and the practical bound starts to degrade.}
  \label{fig:tofino-cap}
\end{figure}
\end{enumerate}

\subsection{Sessions Maintained by Routers}
\begin{itemize}
\item \textbf{How many sessions a router must maintain:} To get an estimate of the number of sessions to expect on routers on the Internet, we have analyzed CAIDA dataset from 2016\footnote{\url{http://www.caida.org/data/passive/passive_2016_dataset.xml}}. The dataset is published every 3 months, consisting of one-hour long anonymized packet headers  
 for an Internet backbone link, operating at 10 Gbps.
Our analysis shows that  on average there are 6,206,033
concurrent source IP and source TCP port tuples for the duration of an hour.

\item \textbf{How much memory do these sessions require:} For each IPv4 session, instead of using all possible combinations of source IP address and port as table keys,  \protocolsp uses a hash function to reduce the addressing space of the memory 
used to store session mappings. Truncating the output of the hash function to 24 bits, yields more than 16 million possible entries (an upper bound for the number of concurrent sessions we observed in the CAIDA dataset). For each entry in the match-action table, \protocolsp stores 4 bytes of original IP address and 2 bytes of original source port. Thus, the total memory required for forward mapping is 96 MB. Similarly for reverse direction the same amount of memory is required, adding up to 192 MB for all entries.

\item \textbf{How many public IPv4 addresses are needed:}  To compute the number of unique IPv4 address needed to map all sessions in the memory, we first computed the maximum number of concurrent TCP sessions observed in the data as an estimate for the number of sessions to be stored. Dividing this number by 65,536, i.e., the total number of ports available shows that we need about 100 different public facing IP addresses for this router to run our protocol with aging out occurring every hour.
Note that the estimates from these traces are conservative estimates, since most routers will see fewer sessions.

\end{itemize}

\subsection{Size of Anonymity Set}
Previous network-level anonymity proposals only considered number of IPv4 addresses assigned to an AS~\cite{lap, hornet} or IP prefix size distribution~\cite{enlistingisp} for quantifying anonymity. This approach does not take into consideration the number of \emph{active IP addresses} and overestimates the AS sizes.
 We, on the other hand, take an active measurement approach to address this limitation. To study anonymity size associated with each service provider on the Internet, we  count the number of active clients on each service provider. 
To this end, we consider the number of active IPv4 addresses as a \emph{lower bound} of active hosts in each service provider. Previous research shows that a combination of active scans on ZMap~\cite{zmap} platform and data observed by Content Distribution Network (CDN) vantage points can  provide an estimate for the number of active IPv4 addresses~\cite{beyondcounting}. However, there is no breakdown by service provider in previous research. For our purposes, we used Censys~\cite{censys}, a platform that captures and processes Internet-wide scanning datasets (using ZMap-based scans) of multiple protocols, such as ICMP reachability, identifying DNS resolvers and HTTP(S) server.  Using Censys we queried the aggregated database of scans for 10 days in July 2018 and on average found  over 156 million active IPv4 addresses that responded to Censys queries. We then found the AS numbers associated with these IP addresses. 

\begin{figure}%
    \centering
    \subfloat[]{
	    \includegraphics[width=0.45\textwidth]{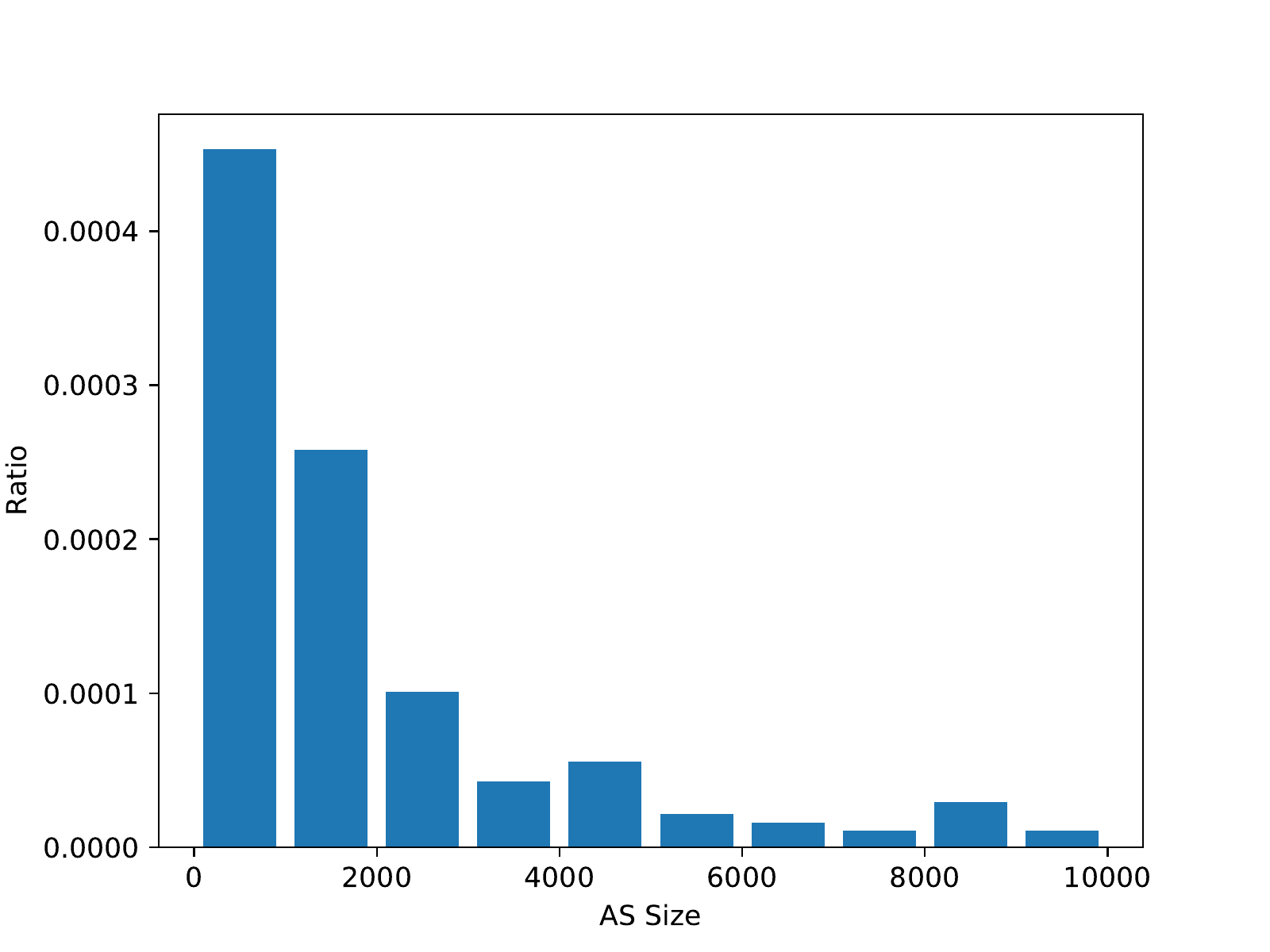} 
    	\label{fig:censys-data1}
    }\qquad
    \subfloat[]{
	    \includegraphics[width=0.45\textwidth]{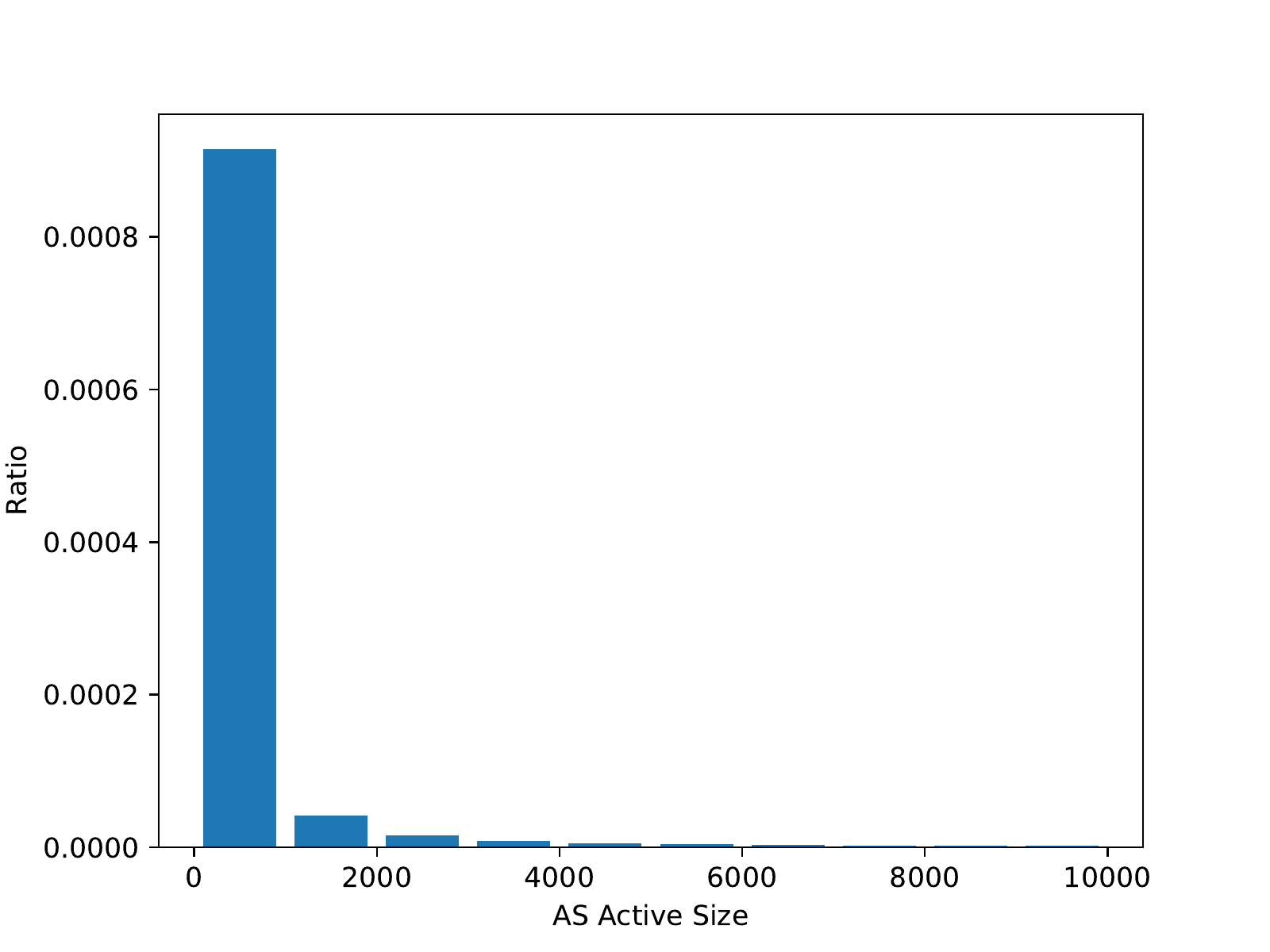} 
    	\label{fig:censys-data2}
    }
    \qquad
    \subfloat[]{
	    \includegraphics[width=0.48\textwidth]{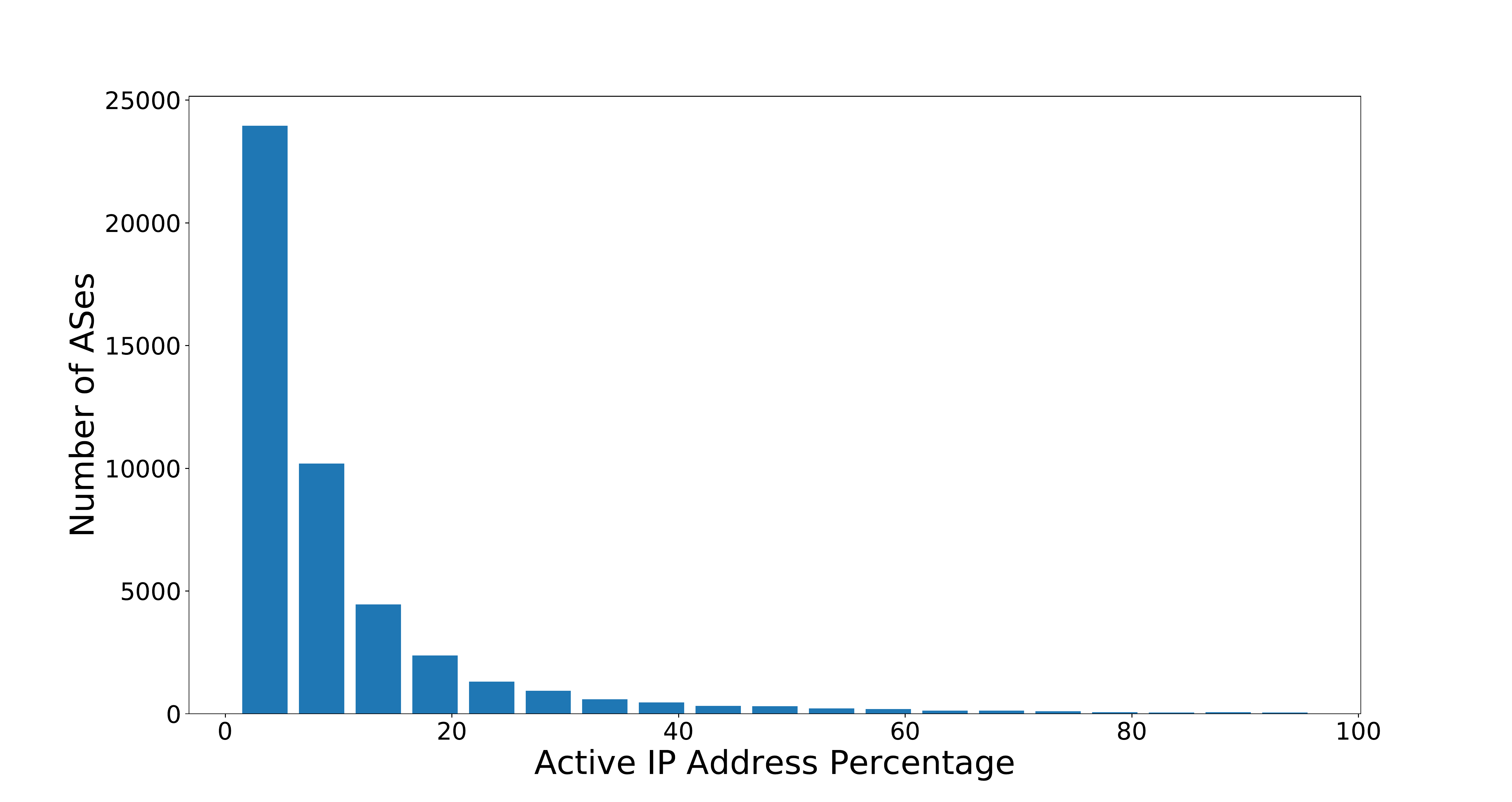} 
    	\label{fig:censys-data3}
    }
    \caption{ The histogram of the total number of IPv4 addresses assigned to each AS is shown in Fig.~\ref{fig:censys-data1} followed by the histogram of the number of active IPv4 addresses per AS, shown in Fig.~\ref{fig:censys-data2}. Note that the horizontal axis is logarithmic. Finally, the percentage of active IPv4 addresses over the number of IPv4 addresses allocated for the AS is shown in Fig.~\ref{fig:censys-data3}}%
    \label{fig:censys-data}%
\end{figure}

Fig.~\ref{fig:censys-data1} shows that the total number of IPv4 addresses assigned to each AS is on average 50532 $\pm$ 989326
Also, as shown in Fig.~\ref{fig:censys-data2}, the number of active IPv4 addresses per AS is on average 2448 $\pm$ 50794. We also compared the percentage of active IPv4 addresses over total number of IPv4 address assigned to a single AS in Fig.~\ref{fig:censys-data3}, with an average value of 8\%.  While this is a lower bound, Using the data from measurement studies that showed 25$\%$ of hosts on the Internet responded to ping request (see section 5 of ~\cite{gong2012website}), we estimate that 32$\%$ of assigned IPv4 addresses are active.
This indicates that there is adequate vacant IP addresses that can be utilized for \protocolsp IP address pool.

\begin{figure*}[h]%
    \centering
    \subfloat[]{
	    \includegraphics[width=0.4\textwidth]{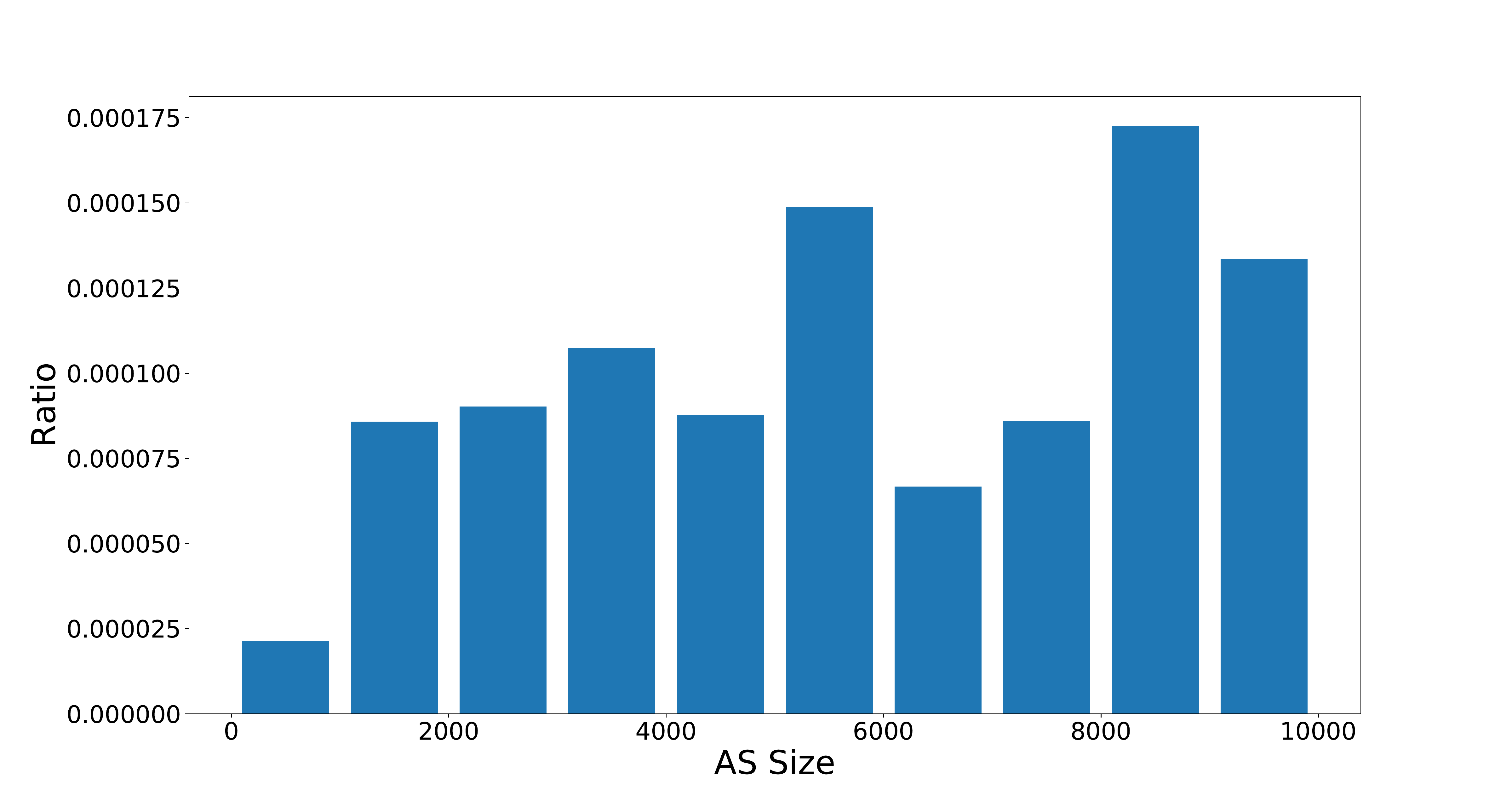} 
    	\label{fig:censys-data1-2step}
    }
    \qquad
    \subfloat[]{
	    \includegraphics[width=0.4\textwidth]{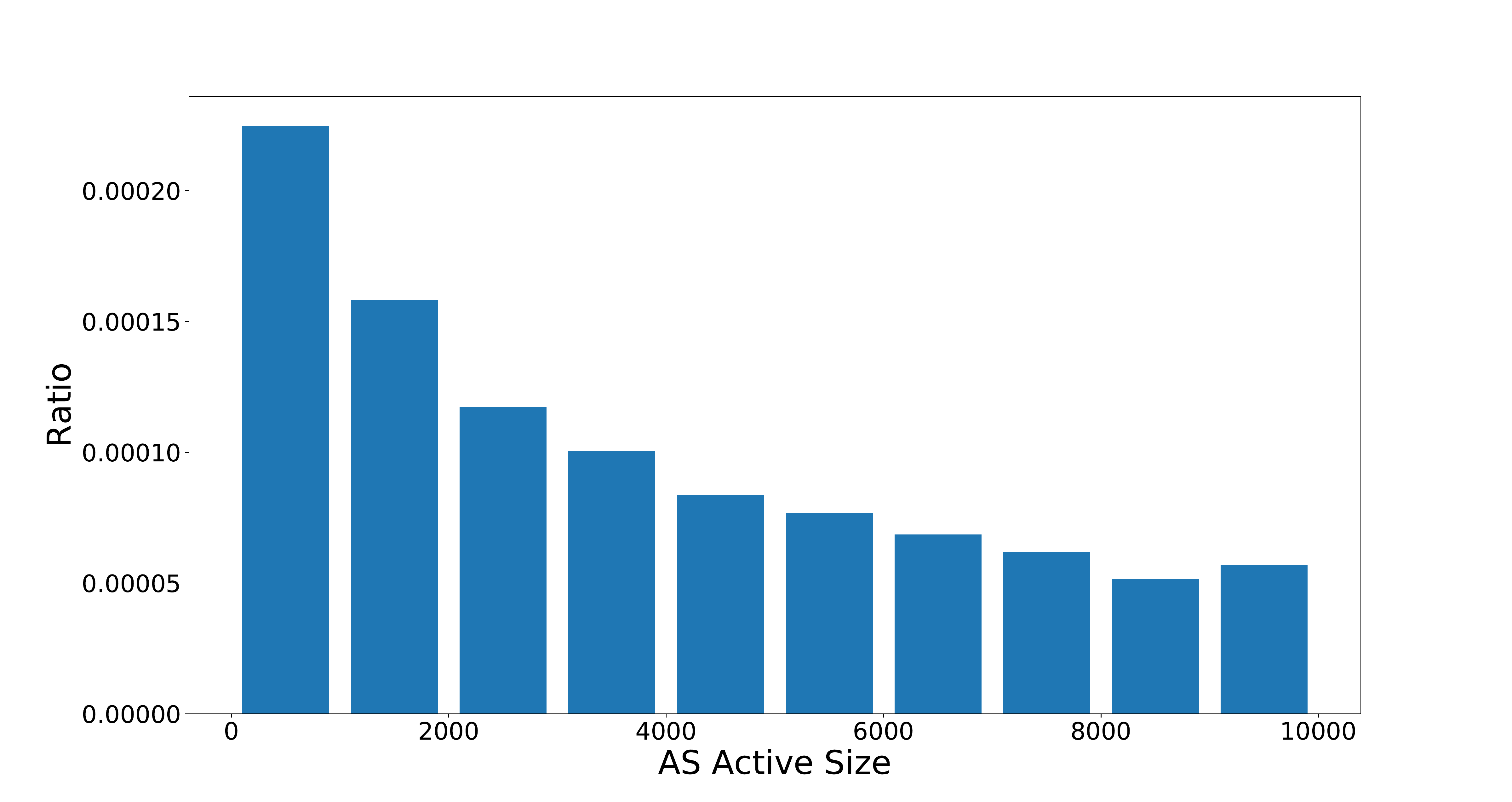} 
    	\label{fig:censys-data2-2step}
    }
    \caption{ The number of total IPv4 addresses assigned to  each two consecutive ASes is shown in \ref{fig:censys-data2-2step} and the number of active IPv4 addresses  in  each two consecutive ASes  is shown in  \ref{fig:censys-data2-2step}. }%
    \label{fig:censys-data}%
\end{figure*}

\hedit{
Finally, To understand how the anonymity set size increases with more ASes implementing \protocol, we assumed that the first two ASes implement our protocol and compute the total anonymity set. For this purpose, we chose Alexa 1000  top domains and their corresponding AS numbers and chose top 1000 ASes based on CAIDA AS ranking\footnote{http://as-rank.caida.org/about}. We then used CAIDA AS relationship database\footnote{http://www.caida.org/data/as-relationships/}  to infer an end-to-end path for sessions originating from the top ASes (source) to the top domains (destination). Using this inferred path, we computed the sum of the total number of IPv4 addresses assigned to the first two ASes in Fig.~\ref{fig:censys-data1-2step} (140989 $\pm$ 142523) and the total number of active IPv4 addresses in Fig.\ref{fig:censys-data2-2step} (average of 3266 $\pm$ 2921). These graphs indicate that the total number of IP addresses assigned and the total number of IP addresses that are active  can increase by a factor of 2.79 and 1.3 respectively, if two consecutive ASes implement \protocol.}

\hcomment{compare this result to previous result.}

\subsection{Distance Distribution for TTL Obfuscation}
\label{eval:ttl}

\begin{figure*}%
    \centering
    \subfloat[]{
      \includegraphics[width=0.4\textwidth]{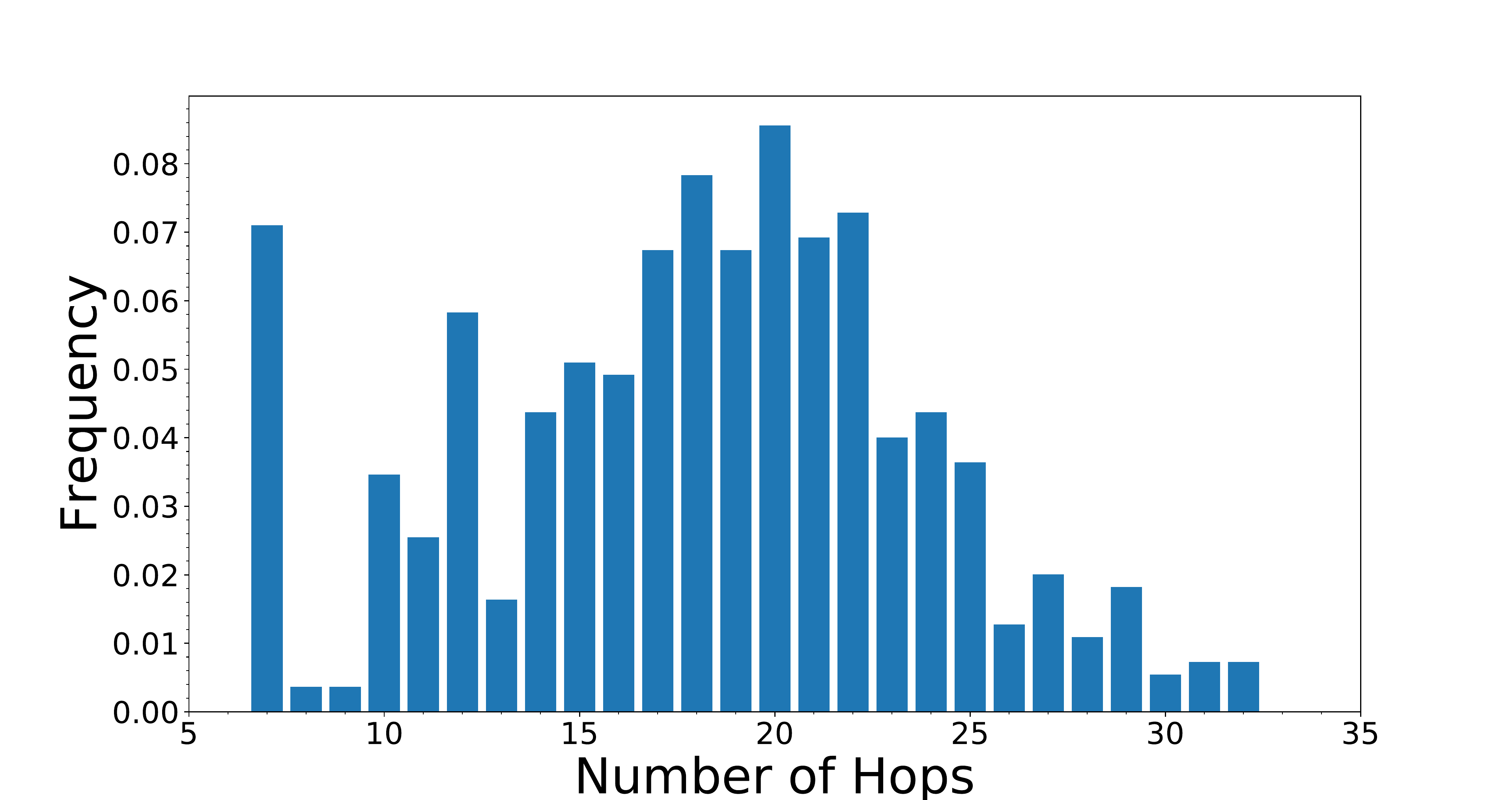} 
      \label{fig:ripe-data1}
    }
    \qquad
    \subfloat[]{
      \includegraphics[width=0.4\textwidth]{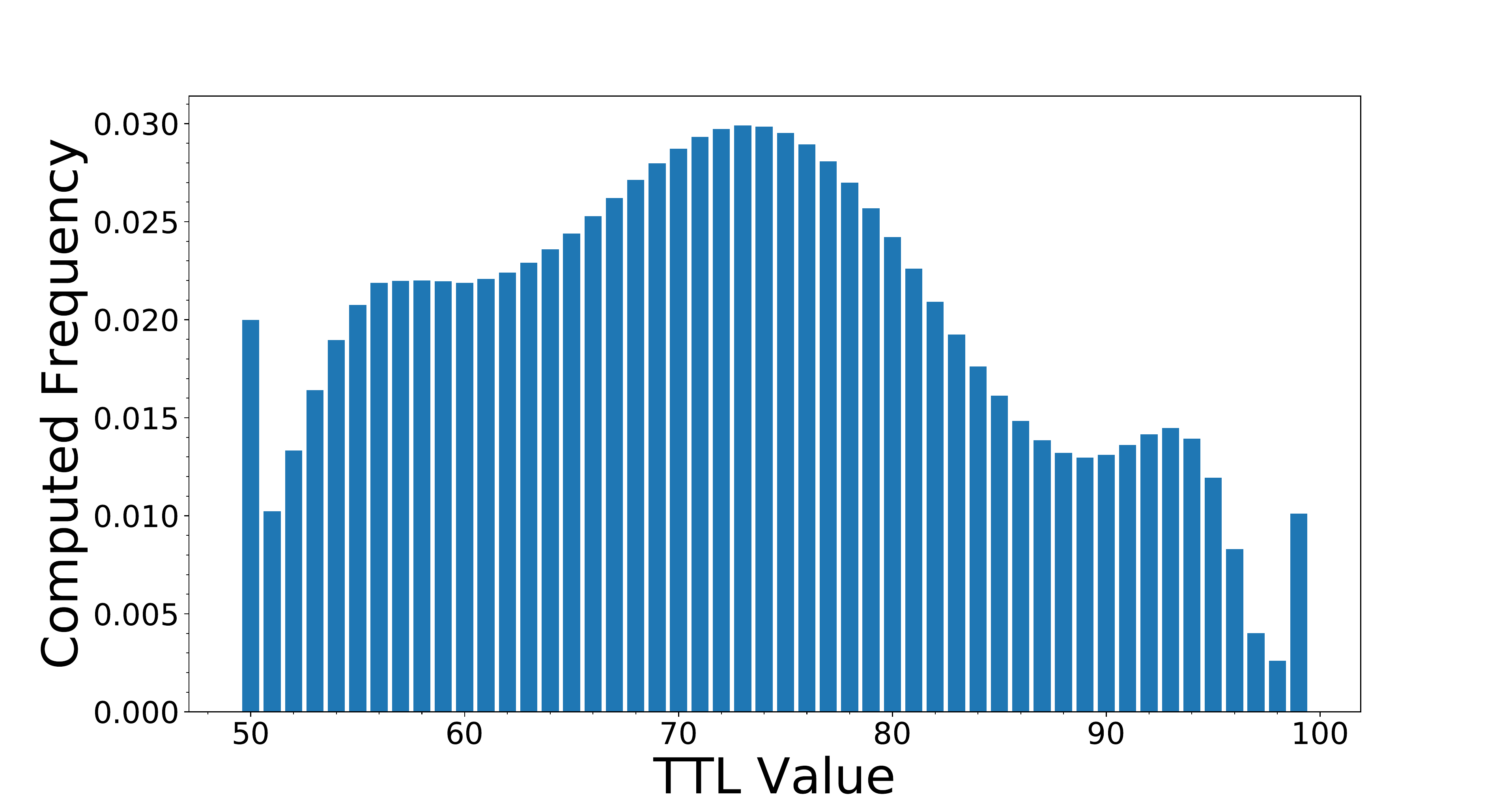} 
      \label{fig:ripe-data2}
    }
    \caption{ Hop distance distribution from vantage points for country Japan is shown in Fig.~\ref{fig:ripe-data1}. The optimized TTL distribution computed based on this distribution is shown in Fig.~\ref{fig:ripe-data2}. }.%
    \label{fig:censys-data}%
\end{figure*}

In Section~\ref{hide_path_info}, we used a distribution to randomize the initial TTL values used by \protocol. In order to estimate that distribution, we first approximated the hop distance to Alexa top 1000 domains. We used RIPE Atlas traceroutes\footnote{https://atlas.ripe.net}, a global network that enables customized active traceroute scans. We collect traceroute data from  3 different vantage points within each country (repeated 10 times) then computed empirical frequency of distance shown in Fig.~\ref{fig:ripe-data1}.
Then, using the equation ~\ref{min_delta_eq} from Section~\ref{hide_path_info}, we found initial TTL distribution by solving the  minimal mutual information equation (Fig.~\ref{fig:ripe-data2}).

\section{Related Work}
\label{sec:related-work}
Network-level anonymity solutions can be broadly categorized into two different classes: 1) enterprise 
and local-area networks, and 2) wide-area networks. 

\textbf{Enterprise and local networks}: In this category, the scope of the anonymity solution is a single enterprise, local network or autonomous system, where all the networking infrastructure is  controlled by a centralized authority. For instance, PHEAR~\cite{noPhear}  replaces identifying information in packet headers with flat tags, which are securely communicated to individual hosts using a centralized  SDN controller. However, this solution requires client side modification. iTap~\cite{itap} takes a similar approach and eliminates the need for a client side software patch.

\textbf{Wide-area network:}
\protocolsp operates in the wide-area network model, thus the following contenders bear more similarity to our design. The Address Hiding Protocol (AHP)~\cite{enlistingisp} proposes a NAT-like design, which aims to accommodate forensics needs by making flow associations reversible, using a cryptographic solution. However, their solution lacks session unlinkability, does not consider legacy network compatibility and makes it difficult for clients in enlisted ISPs to host services. Tor instead of IP~\cite{torinsteadofip} proposed replacing BGP speaking routers with onion routers and rendezvous mailboxes. While embedding onion routing into the network layer would provide both sender and receiver anonymity, the approach demands expensive operations at line-rate and is also based on source-controlled routing.

Lightweight Anonymity and Privacy (LAP)~\cite{lap} aims to protect end-host's identity from servers. In LAP, the first AS is trusted, similar to the PANEL design. However, forwarding information is carried in each packet, with the potential for packet headers to grow arbitrarily large. Also, each AS has to encrypt and verify its own forwarding information as packets pass through the network, performing a symmetric key operation per packet at each hop, as opposed to our solution that only requires a table look-up for packet forwarding.
Dovetail~\cite{dovetail} extends LAP to provide sender-receiver anonymity, with the help of a third party node called \emph{matchmaker}. A client using Dovetail forwards its traffic to the matchmaker and notifies the matchmaker node about the final destination, using public key encryption. Then the sender chooses the final path among a number of paths proposed by the matchmaker. Both LAP and Dovetail leak the location of individual ASes in an end-to-end path between sender and receiver. PHI~\cite{phi}, aims to solve this problem  with a randomized path-segment positioning algorithm.

HORNET~\cite{hornet} offers security and privacy on par with Tor instead of IP, providing an onion encryption with every hop between the two ends of a session.
Moreover, forward and return paths can be distinct in HORNET and information leakage about the length of the path and the location of each AS on the path is minimized. 
However, HORNET assumes that an end-to-end path is known to the sender, where the sender establishes a key for each individual session with every forwarding node on the path, using public key operations. 
Also, similar to Dovetail and PHI, HORNET lacks a clear partial deployment plan and is a source-controlled routing solution, which makes it difficult to deploy in the current Internet. Finally, \hedit{TARANET~\cite{taranet}, improves HORNET, 
adding traffic analysis resilience through a traffic normalization technique called end-to-end traffic shaping. While TARANET has strong anonymity guarantees, its bandwidth and latency overhead are higher than other network-level anonymity proposal.} Table~\ref{tbl:rel-work} summaries the differences between wide-area anonymity solutions.

\begin{table*}[t]
  \centering
  \resizebox{0.65\textwidth}{!}{%
\begin{tabular}{r|ccccccccccc}
&
\rot{Hardware Ready} &
\rot{Partial Deployment} &
\rot{Legacy Network Compatibility} &
\rot{Hop-by-Hop Routing} &
\rot{Sender/Receiver Anonymity} &
\rot{No Packet-Carrying State} &
\rot{No Path Information Leakage} &
\rot{Untrusted First Domain}

    \\ \hline
    
LAP~\cite{lap}		 & \bad{\Circle}	& \mediocre{\LEFTcircle}		& \bad{\Circle}		&   \good{\CIRCLE}		& \bad{\Circle} & \bad{\Circle}	& \bad{\Circle}	& \bad{\Circle}			\\

HORNET~\cite{hornet}			 & \bad{\Circle}& \bad{\Circle}		& \bad{\Circle}		& \bad{\Circle}	 & \good{\CIRCLE} &  \bad{\Circle}	& \good{\CIRCLE}		& \good{\CIRCLE}	 \\

TARANET~\cite{taranet}			 & \bad{\Circle}& \mediocre{\LEFTcircle}	& \bad{\Circle}		& \bad{\Circle}	 & \good{\CIRCLE} &  \bad{\Circle}	& \good{\CIRCLE}		& \good{\CIRCLE}	 \\

Dovetail~\cite{dovetail}		 & \bad{\Circle}& \mediocre{\LEFTcircle}			& \bad{\Circle}		& \bad{\Circle}	& \good{\CIRCLE} & \bad{\Circle}	& \good{\CIRCLE}		&\good{\CIRCLE}		\\


PHI~\cite{phi}			& \bad{\Circle}& \bad{\Circle}		& \bad{\Circle}		& \good{\CIRCLE}	 &\good{\CIRCLE}	& \bad{\Circle}	& \good{\CIRCLE}	& \bad{\Circle}	\\

Tor Instead of IP~\cite{torinsteadofip}	& \bad{\Circle}& \bad{\Circle}		& \bad{\Circle}		& \bad{\Circle}	& \good{\CIRCLE}	& \bad{\Circle}	& \bad{\Circle}	& \good{\CIRCLE}	\\

Enlisting ISPs~\cite{enlistingisp}			 & \bad{\Circle}& \mediocre{\LEFTcircle}	 	& \good{\CIRCLE}	&  \good{\CIRCLE}	 & \bad{\Circle} & \mediocre{\LEFTcircle}	& \bad{\Circle}		&  \bad{\Circle}		\\\

\protocol			 & \good{\CIRCLE} 	& \good{\CIRCLE} 	& \good{\CIRCLE}	&  \good{\CIRCLE}	&  \bad{\Circle}	& \good{\CIRCLE}	& \good{\CIRCLE}	& \bad{\Circle}		\\
\hline
\end{tabular}
}
\caption{Related work comparison. Symbols \good{\CIRCLE} , \mediocre{\LEFTcircle} ,\bad{\Circle} indicate full, partial  and no implementation in the solution respectively.}
\label{tbl:rel-work}
\end{table*}

\section{Discussion}
\label{sec:disc}

In this section, we discuss \protocol's limitations,  extensions for header and payload encryption,  the challenges posed by IP address spoofing, implications for participating ASes, and comparison to 
carrier-grade NATs. 

\subsection{Limitations}
\label{tag-placement}
\label{hard-imp-sess-est}

\hcomment{Return sender anonymity:}

While the design of \protocolsp provides 
several attractive properties including 
partial deployment and being based on hardware switching, it does have several 
limitations, which we discuss next.

\begin{itemize}

\item\hedit{ \textbf{First Round Time Trip Delay}: Recall that session initiation in \protocolsp engages an external CPU/Controller which incurs an extra overhead as shown in Section~\ref{sec:valid-perf}. The P4 specification~\cite{p4spec} allows for external hardware to implement customized functions, such as our session initiation algorithm and performs the operations at line-rate. Platforms such as  NetFPGA\footnote{https://github.com/NetFPGA/P4-NetFPGA-public} can be used for a full  hardware implementation of \protocol, which will reduce the latency overhead in the first round trip of the communication session.
}

\hedit{
\item \textbf{First AS Trust}:  \protocolsp is a light-weight anonymity protocol, as mentioned in Section~\ref{threat-model}. This implies that protocol operations are performed by \protocolsp ASes, and the  first AS has to be the point of trust. However, trusting the first AS can lead to privacy concerns. Users requiring stronger privacy can still benefit from the composition of \protocolsp with the Tor network, as discussed in Section~\ref{sec:composition}.
}

\item \textbf{Tag Placement:}
The number of sessions a \protocolsp router can serve is bounded by the number of bits in packet headers allocated for session tags. Relying only on the port number and lower bits of the IP address might be insufficient for router with large forwarding tables, since it increases the chances of collision. 
To extend tags for IPv4 sessions, we can leverage the literature in IP packet marking, where the goal is to embed information in packets to be able to identify them later in upstream hops~\cite{packetmarking2}. 
 IPv6, on the other hand, provides a much larger address space for this purpose.


\item \textbf{TCP Fingerprintability}: The variance between  TCP implementations on different operating systems allows an adversary to fingerprint clients' operating systems~\cite{TCP/IP-Stack-Fingerprinting}. To obfuscate subtle differences in TCP state transition or advertised TCP window sizes, a \protocolsp segment must employ network-layer security protocols such as IPSec. Thus, TCP fingerprinting remains a limitation of \protocolsp and segments may choose to use more heavy-weight network-level anonymity solutions that also protect against timing analysis, such as TARANET~\cite{taranet} with higher overhead.



\end{itemize}

\subsection{Header and Payload Encryption}

We recognize that the advances in the field of networking will allow us to expand \protocol's feature set. For instance, we discuss how to use line-rate encryption to remove state from landmark routers.

\textbf{Routing Encryption:} Encrypting routing information is rather expensive to perform at line-rate and it is not currently supported by many routers. 
However, if routers are equipped with a cryptographic co-processors, our design can be extended to encrypt local routing decisions and replace tags generated using PRNGs in packet headers, as discussed in Section~\ref{tag-generation}. Furthermore, currently the existence of a particular tag in upstream and downstream packets indicates that they belong to the same end-to-end session. If public key encryption is possible at packet level, tags in packet headers can be encrypted using routers' public keys to provide unlinkability between packets in different directions for a session.

\textbf{Payload Encryption:} \protocolsp neither modifies the payload of packets nor does it perform onion encryption. We believe that payload encryption should be addressed in higher layers, such as IPSec or HTTPS~\cite{halfTrafficEncrypted}. 
  Nevertheless, payload encryption, if possible at line-rate, can help improve anonymity and unlinkability.

\subsection{Address Spoofing and Abuse Mitigation}

\textbf{Address Spoofing}: IP address spoofing entails an attacker using an IP address not belonging to its address space as the source of spoofed packets. \protocolsp routers store per-session state in forward and reverse tables, and an adversary can exhaust the memory on the router. A session is determined by the source IP address and its tag (consider the port number as part of the tag). 
This enables an adversary to have another attack vector namely \emph{tag spoofing},  in which the adversary can change tags while keeping IP address persistent to amplify the attack.


A general solution for tag spoofing is to rate limit tag generation in the first landmark AS.  For orchestrated DDoS attacks, we envision filtering based on the characteristics of the attack~\cite{paxsonDDoS}.
There are also a number of filtering approaches based on characteristics of sessions, such as congestion feedback systems based on destination~\cite{netFence} or capability-based solutions~\cite{vta,portcullis,passport}.

\textbf{Abuse Mitigation}:
Since \protocolsp removes source information from sessions, a server under Denial of Service (DoS) attack does not have any means to filter out connections based on their source address. Thus, we envision a STOP mechanism by which an AS hosting a server can request the first landmark (whose identity must be kept secret due to iterative path mixing) on the path of a session to block certain sessions, on behalf of the victim of the DoS attack. \hedit{
In the absence of return address for a \protocolsp packet, an AS can use the same packet header, to send this STOP request back to the first \protocol segment.
To prove that the STOP packet is generated by the end host's AS, we assume there exist a PKI scheme for Autonomous Systems similar to S-BGP~\cite{sbgp}.
Using its private key, the last AS signs a STOP message and sends it with the same session tag to the first landmark, which in turn will terminate an abusive sender, an approach similar to StopIt~\cite{stopit} and AITF~\cite{atif}.
}

\subsection{Implications for Participating ASes}
\label{impl-ases}

One incentive for ASes to deploy \protocolsp would be to pitch privacy as a differentiating factor, similar to how today's web browsers are pitching privacy to attract users~\cite{firefox-crosssite}. Below, we discuss the cost implications for participating ASes. 


\textbf{Equipment Costs}: Equipment costs scale with the number of sessions an AS has to handle. An AS can choose to cap the number of external sessions it anonymizes. This presents an interesting trade-off between the anonymity set size and economic cost, since larger number of supported sessions lead to a larger anonymity set size, albeit with a higher cost.

\textbf{Limited IP Addresses}: The challenge of a limited number of publicly available IP addresses, especially for IPv4 addresses, can be addressed by having a \protocolsp segment share a pool of public IP addresses, and using ILNP~\cite{ilnp} style location identifier to forward packets to the appropriate router, where the \protocolsp session mapping is stored.

\textbf{Asymmetric routing costs}: Forwarding return packets in asymmetric routes increases the bill for an AS. However, an AS can lower this extra cost by capping the number of external sessions it anonymizes. We also note that ASes have the strongest incentives to support the privacy of their direct customers. Thus, the connections of these users can be prioritized in this process. 

\textbf{Reputation costs}: An AS can be flagged as untrustworthy for forwarding abusive sessions not originating from that AS. The design of abuse mitigation systems~\cite{torpolice,privacypass} is currently an open question for all anonymous communication systems (even those based on overlay mechanisms, such as the Tor network).

\subsection{Comparison to Carrier-grade NAT}
Historically, network address translation (NAT)~\cite{natRFC} has been widely used to work around the challenge of a limited number of IPv4 addresses. As this problem has become more acute in recent years, Internet service providers have started deploying carrier-grade NAT (CGN), also known as large-scale NAT, across their networks. CGNs allow a small pool of public IPv4 addresses to be shared among many hosts and customers, by placing the NAT logic on a service provider's access network, rather than the subscriber's   edge network. Therefore, CGN shifts the NAT function from the customer premises to the service provider network.

We note that the CGNs themselves provide very basic anonymity properties, by multiplexing multiple hosts and customers via shared IPv4 addresses. The design of \protocolsp indeed shares some similarities with CGNs, and builds upon the basic anonymity properties they provide. However, \protocolsp also differs from CGNs in several important ways. 

First, \protocolsp rewrites source information in a more rigorous and holistic manner than CGNs, including normalizing IP identification field, normalizing TCP initial sequence number, and hiding path information via time-to-live randomization. Furthermore, the mappings between sessions and session identifiers (tags) in \protocolsp are not chosen deterministically, but are chosen in a randomized manner using a PRNG as described in Section~\ref{des:ses-unlink}. In contrast,  such mappings (session outbound ports) in CGNs are deterministic. As shown by prior work, this determinism can be used to link sessions of users, especially in the case of permissive (full cone) NAT, which is quite prevalent in design of CGNs~\cite{cgnStudy}.

Second, \protocolsp uses iterative path mixing to further enhance anonymity, via protocol deployment at multiple \protocolsp segments in an end-to-end communication path 
between a client and server. 

Third, \protocolsp does not act as a multiplexer of internal destinations and thus, does not impose the kind of restriction that classical NATs place on network endpoints behind a NAT. For instance, in our design, a client can use the address of a server located in a landmark segment directly to access it. In fact, we argue that there need to be enough IP addresses allocated to individual hosts. 
Furthermore, \protocolsp explicitly mandates that service providers do not log the mapping between internal endpoint and global tags as discussed in Section~\ref{sec:sys-design}.

%
\protocolsp also shares a number of common challenges with CGNs~\cite{cgnReport}. 
 First, for scalability, both need to maintain device throughput and line-rate packet processing. Moreover, resource management, such as memory used to keep sessions state or IP pool size (number of public facing IPs) is also a key component of design for both technologies. 
 There are a number of best practices around designing CGNs and protocols that interact with them~\cite{cgnReqRFC} and studies have shown the result of these practices in different providers~\cite{cgnStudy}. Some of the techniques for DoS mitigation such as ingress filtering~\cite{ingressFilter}, TCP/UDP timeout are lessons applicable to both \protocolsp and CGNs.

\section{Conclusions}
\label{sec:conc}
In order to address one of the most challenging impediments that network-level anonymity proposals face (namely, \textit{deployability}), we introduced \protocolsp. 

The proposed approach is sufficiently simple to be implemented using primitives available in programmable hardware switches: we prototyped \protocolsp on a real-world Barefoot Tofino switch using P4~\cite{p4} network programming language. \protocolsp offers (1) partial deployability, (2) low-latency, and (3) transparency (i.e., it does not require client-side or server-side modifications).

Our analyses demonstrate that \protocolsp offers a high throughput (96\% of the actual throughput of the switch), low-latency overhead (e.g., 3\% overhead in Skype calls), and satisfactory anonymity set at the cost of a reasonable storage overhead (192 MB).
\bibliographystyle{unsrtnat} 
\bibliography{panel}
\appendix

\section{Appendix A.}
\label{appA}
\subsection{TCP Handshake Latency}
\begin{figure*}[]
   \centering 
	    \includegraphics[trim={20mm 1mm 35mm 2mm},clip,width=\textwidth]{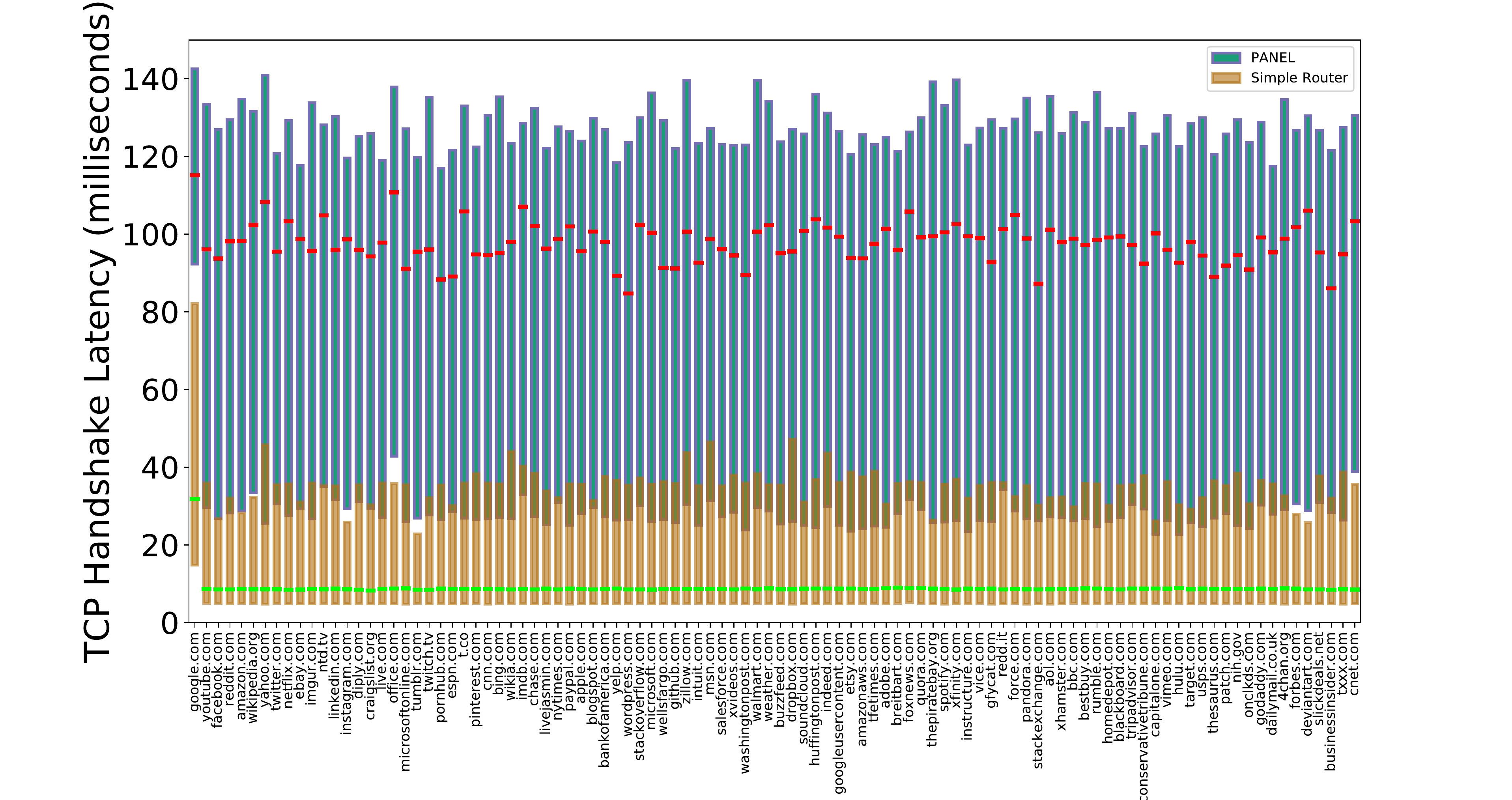} 
  \caption{Comparison of TCP handshake latency for Alexa top 100 domain over \protocolsp and simple router. This graph shows the time to complete a TCP three-way handshake for each domain.}
      	\label{fig:tcp-latency}
\end{figure*}

This next experiment shows that this overhead is mostly introduced in the first packets of each session: we established 1000 TCP sessions to each of the above mentioned websites over TCP port 80 and measured the time it took to complete the TCP handshake. As shown in Fig.~\ref{fig:tcp-latency}, for simple router, the average time to complete the TCP handshake was 18.22 millisecond per session. In contrast, the average time to complete the TCP handshake was 152.16 millisecond per session for \protocol, showing an 8 fold increase in the first round trip. This overhead can be reduced substantially once the \protocolsp \emph{session establishment} logic is moved to a full hardware implementation, as opposed to our current python based local agent (See Section~\ref{hard-imp-sess-est}).

\end{document}